\documentclass[lineno]{jfm}

\usepackage{graphicx}
\usepackage{subfigure}
\usepackage{epstopdf,epsfig}
\usepackage{newtxtext}
\usepackage{newtxmath}
\usepackage{natbib}
\usepackage{hyperref}
\hypersetup{
    colorlinks = true,
    urlcolor   = blue,
    citecolor  = blue,
}

\newcommand{\RomanNumeralCaps}[1]
\linenumbers
\usepackage{booktabs}

\newcommand{\Ue}{\bar{U}}
\newcommand{\dd}{\mathrm{d}}
\newcommand{\ii}{\mathrm{i}}
\newcommand{\ee}{\mathrm{e}}
\newcommand{\Real}{\mathrm{Re}}

\usepackage[T1]{fontenc}
\usepackage{amsmath}

\usepackage{tikz}

\title[Weak internal waves over a ridge]{Weakly nonlinear internal waves by tidal flow
over a ridge in a shear current}

\author[X. Huang]{Xun Huang\thanks{Email address for correspondence: huangxun@pku.edu.cn}
  }

\affiliation{State Key Laboratory of Turbulence and Complex Systems, School of Mechanics and Engineering Science, Peking University, Beijing, China}

\begin{document}
\fulloutfootnote
\maketitle

\begin{abstract}
\noindent We extend Thorpe's successive-approximation expansion to the forced, tide-locked internal-wave generation problem of Lamb and Dunphy, where a barotropic tide over a ridge radiates a discrete spectrum of Taylor--Goldstein eigenmodes in a steady shear current. At second order in topographic steepness, each mode forces a bound second harmonic via diagonal mode-pair kernels derived in closed form; both kernels vanish for uniform flow, recovering Thorpe's limit. At third order, solvability of the resonant fundamental yields a nonlinear wavenumber correction—the forced counterpart of Thorpe's phase-speed shift, with tidal frequency fixed. Computations at the parameters of Lamb and Dunphy show that the corrections are small but systematic, with the bound harmonic distorting the displacement profile modestly and concentrated in the surface shear layer for downstream modes. The corrections peak at modes 2--3, grow quadratically with ridge height and current strength, and amount to a nonlinear slow-down of every mode, explaining from within the theory why the linear discrete-spectrum model agreed so well with fully nonlinear simulations. A pycnocline case with stratification co-located with the shear layer amplifies the nonlinear corrections substantially --- the peak wavenumber correction grows by a factor of about $67$ and the waveform distortion reaches $8\%$ --- showing that the uniform-stratification verdict is not generic and that the corrections are controlled by where $N^2$ sits relative to the shear. The machinery developed provides a path from weakly nonlinear corrections toward a quantitative finite-amplitude theory of topographic internal tides.
\end{abstract}

\section{Introduction}
\label{sec:introduction}

The generation of internal waves by barotropic tidal flow over seafloor topography represents a cornerstone of physical oceanography, functioning both as the primary energy source for the ocean's internal wave field and as a critical driver of diapycnal mixing and turbulence in stratified fluids \citep{Lamb2014}. Accurately simulating the full development of these waves, however, involves a prohibitively demanding numerical effort due to the wide spectrum of length scales that must be resolved \citep{KlemaVenayagamoorthy2024, Sampatirao2026}. Despite recent substantial advances, including the three-dimensional modelling of internal tide generation over isolated seamounts in a rotating ocean \citep{LeDizes2025}, 
idealised model approaches that capture the salient features of these representative processes remain a powerful approach. They not only provide deepened physical insight but also offer a practical means to rapidly quantify the relative significance of the various governing mechanisms  \citep{Howland2021,Kav2025,Sampatirao2026}.

\begin{figure}
\begin{center}
\begin{minipage}{130mm}
\begin{center}
    \subfigure[ ]
              {\includegraphics[width=120mm]{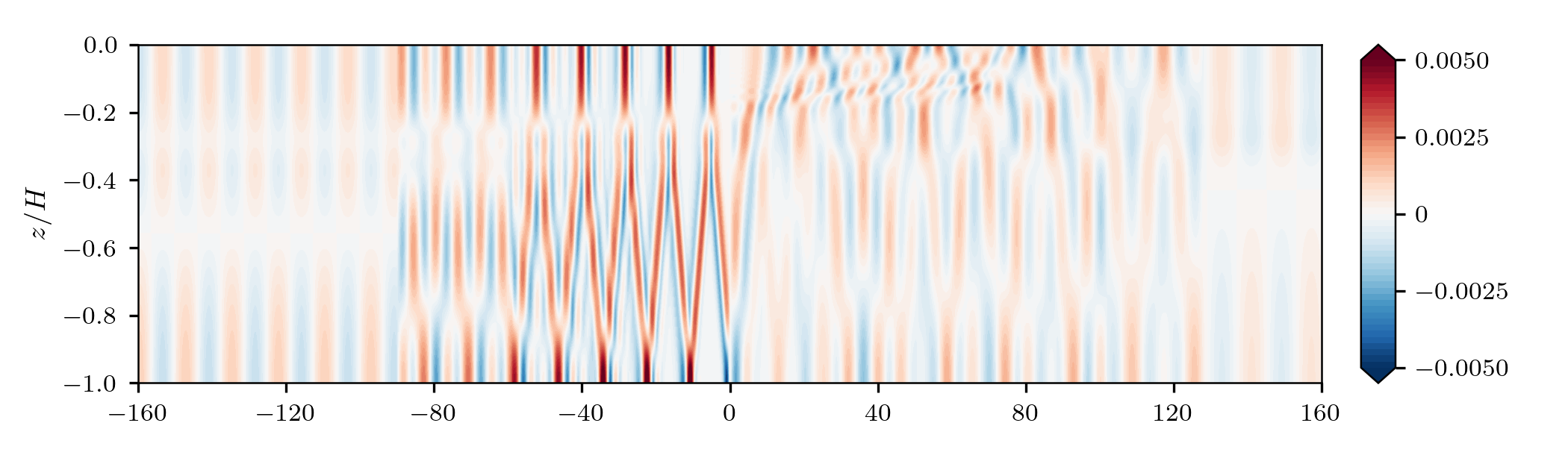}}\\
    \subfigure[ ]
              {\includegraphics[width=120mm]{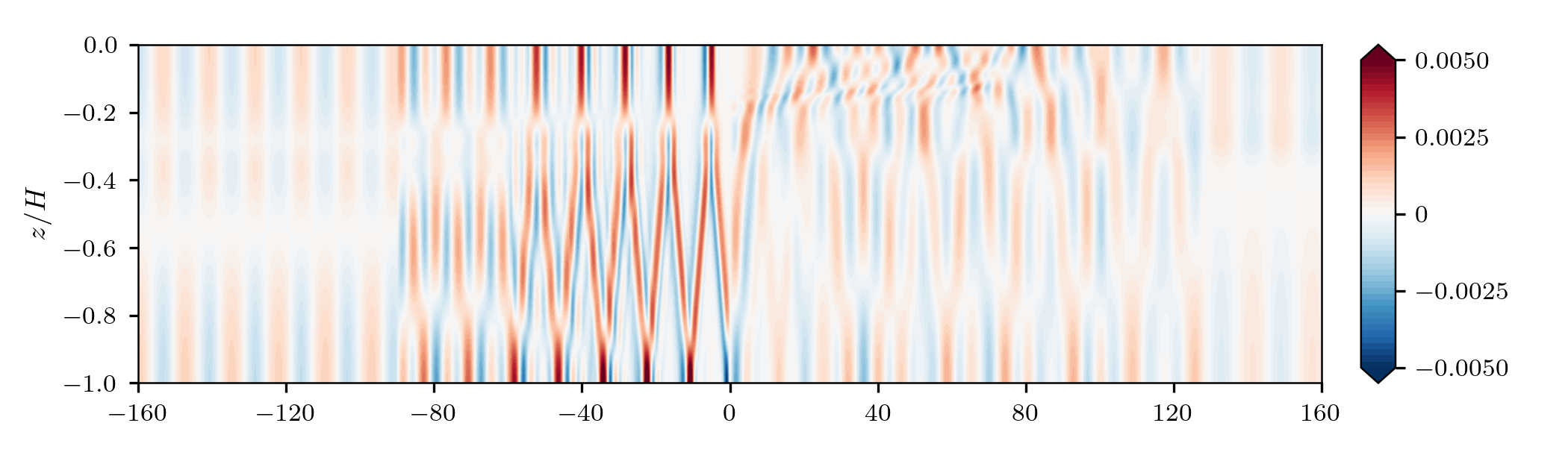}}\\
      \subfigure[ ]
              {\includegraphics[width=120mm]{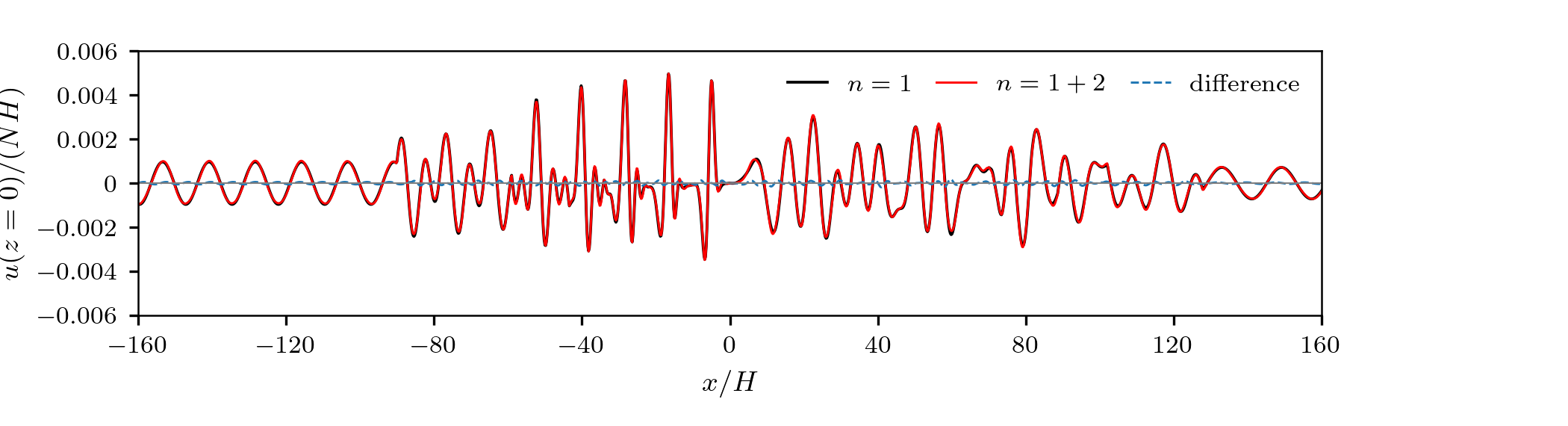}}            
  \caption{\label{f:fig1} Comparison of wave-induced dimensionless horizontal current fields $U/(NH)$  at $t = 15\,T$ from  \citet{LambDunphy2018}'s linear model with (a) $n=1$ and (b) $n=1,2$, where $N$ is the constant buoyanancy frequency (set to $10^{-3}$), $H$ the deep see depth ($5000\,$m), $T$ is associated with the principal lunar semidiurnal tide (12.42\,h), and $n$ refers to waves of frequency $n\omega_0$ with $\omega_0=2\pi/T$. For both $n$ harmonics, 24 discrete vertical modes are calculated to consistute the whole field. 
  Other set-ups are the same as those in figure~1 in \citet{LambDunphy2018}. For clarity of the figures, the bottom ridge is not shown here but will be explained in the next figure. Panel (c) compares the differences between panels (a) and (b) at the surface that shows the negligible second harmonic for the current set-up.    }
\end{center}
\end{minipage}
\end{center}
\end{figure}
%
\citet{Bell1975} laid the theoretical foundation for internal wave generation by bottom topography, demonstrating that tidal flow over a two-dimensional obstacle in a uniformly stratified, unbounded fluid radiates internal waves at the fundamental frequency and at all harmonics below the buoyancy frequency. 
\citet{Khatiwala2003} extended the model to finite depth, decomposing the response into vertical normal modes. As shown in figure~\ref{f:fig1},   
 \citet{LambDunphy2018} (hereafter LD18 in figures) introduced a crucial new physical element -- a steady, surface-trapped background current -- thereby capturing progressively more realistic and complicated physical configurations. 
 In  \citet{LambDunphy2018}'s theory the baroclinic response is decomposed into tidal harmonics whose vertical structure is governed by the Taylor--Goldstein
 equation; the radiated field is dominated by a discrete spectrum
of eigenmodes whose phase speeds are Doppler-shifted and whose
upstream/downstream symmetry is broken by the surface-trapped current, while the total
conversion remains essentially insensitive to it.  Validated against
fully nonlinear simulations using large eddy simulation, the linear model based on discrete spectrum proved
remarkably accurate.

Two issues were left open by \citet{LambDunphy2018}.  The first is the role of the
continuous spectrum associated with critical-level singularities of the Taylor--Goldstein operator, neglected in the residue evaluation.  This has recently been resolved analytically by \citet{OnukiVenaille2025}, who
showed that the continuous-spectrum contribution to the conversion
budget is small at the fundamental frequency --- although its wave
packets develop algebraic shear growth downstream, an inviscid route
to breaking.  The second open issue is nonlinearity itself, and it
motivates the present work. A physical rationale for the irrelevance of finite-amplitude effects is still required, along with a parametric map delineating the conditions under which these effects first become significant.

That nonlinear corrections eventually matter is beyond doubt. \citet{GayenSarkar2011} found in direct and large eddy simulations of internal-tide generation at a near-critical slope that the second
harmonic carries up to 15\% of the fundamental's energy once the
resonant fetch is long enough, and that turbulence saturates the
conversion at 25--40\% of the linear-theory flux. For the present set-up, however, figure~\ref{f:fig1} reveals that the second harmonic contributes negligibly, so we henceforth direct our focus to higher-order nonlinear effects.
 \citet{KlemaVenayagamoorthy2024} showed that for mode-1 waves
impinging on ridges the reflection and dissipation  are strongly nonlinear in ridge height and wave
Froude number, with non-hydrostatic pressure work contributing up to
half of the energy flux. 

Furthermore,  \citet{BuhlerMuller2007} demonstrated that the linear
internal-tide field can overturn along characteristics radiating from
topographic curvature singularities, and \citet{GuoHolmesCerfon2016}
showed that random small-amplitude topography generically supports
wave attractors along which inviscid linear gradients grow without
bound.  Observationally, internal-wave slope spectra saturate ---
amplitudes stop growing while dissipation varies significantly
\citep{KlymakMoum2007} --- the signature of a breaking-limited,
fundamentally nonlinear regime. Most recently, \citet{Sampatirao2026} proposed a semi-analytical Green function framework that simultaneously models internal wave generation by barotropic forcing and scattering of an incident mode-1 internal wave at bottom topography, revealing that their phase-dependent interference can cause total energy flux to deviate significantly from linear superposition and even lead to energy loss of the incident wave.

In this study, our application of finite-amplitude internal-wave theory to shear flows essentially follows \citet{Thorpe1978}. Expanding a freely propagating internal mode as a perturbation series in the wave steepness $\varepsilon$ (the ratio of vertical displacement amplitude to wavelength), \citet{Thorpe1978} showed: (i) at $O(\varepsilon^2)$, nonlinear self-interaction generates a second-harmonic correction  that distorts the waveform—an effect significantly amplified by shear through the secular growth of the distortion amplitude, especially near critical Richardson numbers; (ii) at $O(\varepsilon^3)$, the Fredholm solvability condition yields an $O(\varepsilon^2)$ correction to the phase speed, arising from the nonlinear modification of the dispersion relation. This work aims to extend that analysis framework from freely propagating wavetrains to internal waves forced by bottom topography.

Related kinematic and convective breaking criteria were developed by \citet{OrlanskiBryan1969} and \citet{Sutherland2001}, and three-dimensional simulations have since clarified what such
criteria can and cannot predict: \citet{Howland2021} found that the
overturning condition (unit wave steepness) is not a qualitatively
significant threshold for the nature of the breakdown, that linear
refraction theory remains a useful guide to where wave energy
accumulates in a shear flow even at large steepness, and --- of
direct relevance here --- identified the extension of \citet{LambDunphy2018}'s
two-dimensional ridge--shear set-up as a natural next step.  Their
simulations further showed that the turbulence triggered by
shear-induced breaking retains a mixing efficiency comparable to
Kelvin--Helmholtz billows in a stratified shear layer, even though the
overturning wave strongly distorts the buoyancy field on which the
billows develop, and that reducing the wave amplitude delays the onset
of the instabilities rather than changing their character.  These
findings bear directly on the present configuration: they identify the
surface shear layer as the seat of wave-energy accumulation and
instability --- precisely where the finite-amplitude corrections
computed below are found to be largest --- and they support using the
weakly nonlinear modal structure to anticipate where the tidally
forced wave field first approaches the finite-amplitude regime. 

On the other hand, wave
propagation through sheared, marginally stable flows is meanwhile well
described by the linear Taylor--Goldstein framework \citep{PhamSarkar2010}, so the
baseline onto which the nonlinear corrections are grafted is secure.
\citet{Thorpe1978}'s analysis, however, applies to a single free mode with an
adjustable frequency --- not to a tide-locked forced problem with a
whole discrete spectrum of modes.  The present paper closes exactly
that gap.

In this work, we embed \citet{Thorpe1978}'s successive-approximation expansion into \citet{LambDunphy2018}'s
forced architecture and answer two questions.  First, {at the
parameters of  \citet{LambDunphy2018}'s own figures~5--6, how large are the second-order
waveform distortion and the third-order nonlinear dispersion, and what are the associated fluid mechanics?}
Second, {how sensitive are these corrections to the structure of
the stratification --- in particular to the oceanically common
situation, e.g.,the  \citet{Thorpe1978}'s pycnocline, in which the buoyancy-frequency maximum
coincides with the shear layer?}  The frequency being
locked to the tidal harmonic $n\omega_0$, the nonlinear degree of
freedom is the modal wavenumber. All
calculations are carried out with the discrete-spectrum code of the
linear theory, extended by the forced Taylor--Goldstein
boundary-value solves and projection integrals
reported below.

The remaining part of this paper is organised as follows.  Section~\ref{sec:problem} states the
problem and reviews the first-order discrete-spectrum solution. Section~
\ref{sec:higher-order} develops the $O(\varepsilon^2)$ and $O(\varepsilon^3)$ theory, including the mode-pair kernels, the
solvability-based wavenumber correction, triad conditions and the
mean-field problem.  Section \ref{sec:results} presents the dispersion
and waveform results, first for the uniform stratification set-up and then for a pycnocline stratification
co-located with the shear layer. Section \ref{sec:outlook} collects the
directions that remain open. Finally, \S\ref{sec:conclusions} concludes the key findings of this work.

\section{Problem formulation and the linear model}
\label{sec:problem}

\subsection{Statement of the problem}

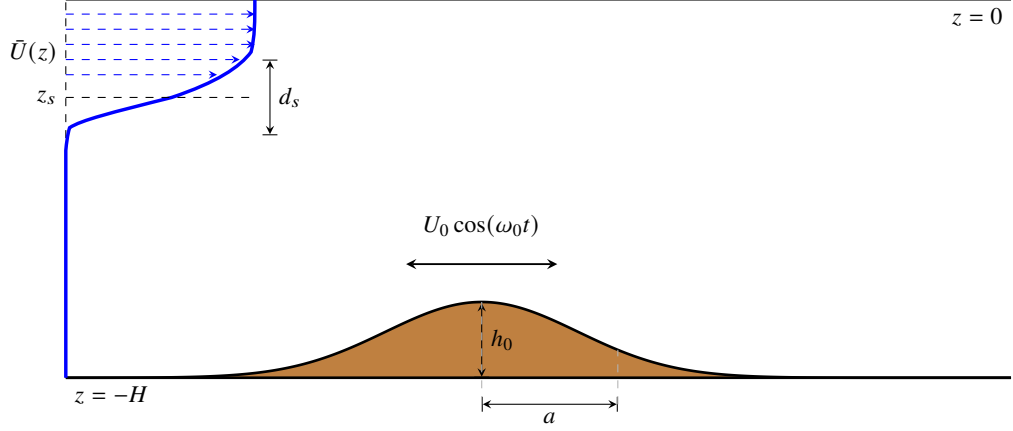
\begin{figure}
\begin{center}
\begin{tikzpicture}[
    >=stealth,
    font=\small,
    currentline/.style={blue,line width=1.3pt},
    stratline/.style={gray!50,thin,densely dotted},
    dimline/.style={|<->|,thin,font=\footnotesize},
    annot/.style={font=\footnotesize},
]
\fill[brown]
    (0,-5) --
    plot[domain=0:12.5,samples=150,smooth]
        (\x,{-5 + 1.0*exp(-(\x-5.5)^2/3.24)}) --
    (12.5,-5) -- cycle;
\draw[black,line width=1.0pt]
    plot[domain=0:12.5,samples=150,smooth]
        (\x,{-5 + 1.0*exp(-(\x-5.5)^2/3.24)});
\draw[black,line width=1.0pt] (0,-5) -- (12.5,-5);

\draw[line width=1.2pt] (0,0) -- (12.5,0);
\node[annot,anchor=north east] at (12.5,0) {$z=0$};

\draw[dimline] (5.5,-5) -- node[annot,anchor=west] {$h_0$}
    (5.5,{-5 + 1.0});
\draw[dimline] (5.5,-5.35) -- node[annot,anchor=north] {$a$} (7.3,-5.35);
\draw[dashed,gray!60] (5.5,-5.35) -- (5.5,-4.0);
\draw[dashed,gray!60] (7.3,-5.35) -- (7.3,{-5+1.0*exp(-(7.3-5.5)^2/3.24)});
\node[annot] at (5.5,-3.0) {$U_0\cos(\omega_0 t)$};
\draw[<->, line width=0.7pt] (4.5,-3.5) -- (6.5,-3.5);
\node[annot,anchor=north west] at (0,-5) {$z=-H$};

\draw[dashed] (0,-5) -- (0,0);
\draw[currentline]
    (2.5, 0)
    .. controls (2.5, -0.3) and (2.5, -0.5) .. (2.45, -0.7)
    .. controls (2.3, -0.9) and (2.0, -1.1) .. (1.4, -1.3)
    .. controls (0.6, -1.5) and (0.2, -1.6) .. (0.05, -1.7)
    .. controls (0.02, -1.8) and (0.01, -1.9) .. (0.0, -2.0)
    -- (0, -5.0);
\draw[->,dashed,blue] (0,-0.2) -- (2.5,-0.2);
\draw[->,dashed,blue] (0,-0.4) -- (2.5,-0.4);
\draw[->,dashed,blue] (0,-0.6) -- (2.5,-0.6);
\draw[->,dashed,blue] (0,-0.8) -- (2.3,-0.8);
\draw[->,dashed,blue] (0,-1.0) -- (2.,-1.0);
\node[annot,anchor=east] at (0,-0.7) {$\bar{U}(z)$};
\draw[dashed,black] (0,-1.3) -- (2.5,-1.3);
\node[annot,anchor=east] at (0,-1.3) {$z_s$};
\draw[dimline] (2.7,-1.8) -- node[annot,anchor=west] {$d_s$} (2.7,-0.8);
\end{tikzpicture}
\caption{Schematic of the problem: a barotropic tide $U_0\cos\omega_0 t$ oscillates over a Gaussian ridge
$h(x)=h_0\ee^{-(x/a)^2}$ in water of depth $H$ with a rigid lid at
$z=0$; a steady surface-trapped current $\Ue(z)$ (\ref{eq:U0}), centred at $z_s$ with shear-layer thickness $d_s$, occupies the upper water column. Two background stratification configurations are examined: a uniform stratification with constant buoyancy frequency as in \citet{LambDunphy2018}, and a non-uniform stratification $N(z)$ designed to mimic a \citet{Thorpe1978}'s pycnocline in which the density gradient is localised within the shear layer.
 }
\label{f:sch1}
\end{center}
\end{figure}
The configuration, shown in figure~\ref{f:sch1}, essentially follows \citet{LambDunphy2018}'s configuration.  A rigid
lid sits at $z=0$; the bottom at $z=-H+h(x)$ carries an isolated
Gaussian ridge
\begin{equation}
h(x) = h_0 \mathrm{e}^{-(x/a)^2},\label{eq:ridge}
\end{equation}
of amplitude $h_0$ and half-width $a$, whose maximum slope is
$s=\sqrt{2/e}\,h_0/a\approx0.86\,h_0/a$.  The steady background
current, trapped against the surface,
\begin{equation}
\bar{U}(z) = \frac{U_s}{4}\left(1 + \tanh\left(\frac{z - z_s}{d_s}\right)\right)^2, \label{eq:U0}
\end{equation}
takes its maximum $U_s$ at the lid and decays through a shear layer
centred at $z_s$ of thickness $d_s$; it vanishes well above the ridge
crest, so the current does not itself feel the topography.  Its
minimum gradient Richardson number,
\begin{equation}
Ri_{\min}=\frac{N^2}{\max(\Ue')^2}
=\left(\frac{27}{16}\right)^2\left(\frac{N d_s}{U_s}\right)^2,
\label{eq:rimin}
\end{equation}
exceeds the Miles--Howard threshold $1/4$ in all cases considered, so
the background state is shear-stable.  The stratification is uniform,
$N=10^{-3}\,\mathrm{s^{-1}}$, as in \citet{LambDunphy2018}'s examples, except in the
\citet{Thorpe1978}-style pycnocline case of \S\ref{sec:pycnocline}, where a
non-uniform $N(z)$ is introduced.  

The barotropic tide
\begin{equation}
U_b(t) = U_0 \cos(\omega_0 t), \label{eq:ub}
\end{equation}
with $\omega_0$ the M$_2$ frequency (period 12.42~h), forces the flow
through the linearised bottom condition
\begin{equation}
w (x, z=-H, t) = U_b(t) \frac{\dd h(x)}{\dd x}, \label{eq:bc}
\end{equation}
with the exact (fully nonlinear) bottom condition expanding
order-by-order as detailed in \S\ref{sec:inhomTG}.  The slope
criticality $\gamma$, the ratio of the maximum bottom slope to the
internal-wave beam slope, is approximately $6.06\,h_0/a$; all cases
computed here are strongly subcritical ($\gamma\approx0.5$), in the
regime where \citet{LambDunphy2018}'s linear theory compared favourably with their large eddy 
simulations.

\subsection{Governing equations}
\label{sec:governing}

Following \citet{Bell1975}, we work in the reference frame comoving with the barotropic tide,
\begin{equation}
\xi = x - \int_0^t U_b(\tau)\dd\tau = x -  \frac{U_0}{\omega_0}\sin\omega_0 t,
\end{equation}
in which the topography is stationary and the response is time-periodic
at the tidal harmonics.  The model is the two-dimensional,
non-hydrostatic, incompressible Euler system under the Boussinesq
approximation used by \citet{Thorpe1978} and \citet{Lamb1994}.  With total
streamfunction $\Psi = \Psi_0(z) + \psi$, $\Psi'_0 = \bar U$,
perturbation velocity $(u, w) = (\Psi_z, -\Psi_\xi)$ and buoyancy
disturbance $b = -g\rho'/\rho_{0}$, the vorticity and buoyancy
equations are
\begin{align}
D_t \nabla^2\psi - \bar U'' \psi_\xi + J(\nabla^2\psi,\psi) &= -b_\xi, \label{eq:me1}\\
D_t b - N^2 \psi_\xi &= -J(b,\psi), \label{eq:me2}
\end{align}
where $D_t = \partial_t + \bar U(z)\partial_\xi$ and
$J(f,g) = f_\xi g_z - f_z g_\xi$.  These are \citet{Thorpe1978}'s equations
(1)--(2) with the disturbance fields separated from the steady
background: his continuity equation for the total density,
decomposed into background plus perturbation, transforms directly
into (\ref{eq:me2}).

Eliminating $b$ by applying $D_t$ to~\eqref{eq:me1} and substituting
$\partial_\xi$ times~\eqref{eq:me2} yields
\begin{equation}
D_t^2 \nabla^2\psi + N^2 \psi_{\xi\xi} - \bar U'' D_t \psi_\xi
= -D_t J(\nabla^2\psi,\psi) + \partial_\xi J(b,\psi).
\label{eq:master}
\end{equation}


Next, all fields are expanded in powers of the topographic steepness
$\varepsilon = h_0/H$:
\begin{eqnarray}
\psi &=& \varepsilon \psi_1 + \varepsilon^2 \psi_2 + \varepsilon^3 \psi_3 + \cdots, \\
b &=& \varepsilon b_1 + \varepsilon^2 b_2 + \varepsilon^3 b_3 + \cdots.
\label{eq:expand}
\end{eqnarray}
The bottom boundary condition is imposed order-by-order via its
Fourier--Bessel spectral decomposition.  At
$O(\varepsilon)$ this yields \citet{LambDunphy2018}'s forced solution (\S\,4 therein); at
$O(\varepsilon^2)$ and $O(\varepsilon^3)$ it provides the
inhomogeneous boundary data $g_2$ and $g_3$ in~\eqref{eq:gdata} below.

\subsection{First-order solution}
\label{sec:first-order-review}

At $O(\varepsilon)$ the right-hand side of~\eqref{eq:master} vanishes.
Seeking solutions $\propto \mathrm{exp}({i(\kappa\xi - \nu t)})$ with
$\nu = n\omega_0$ gives the homogeneous Taylor--Goldstein equation
\begin{equation}
L_n(\kappa)[\phi_n] \equiv \phi_n'' + \left[ \frac{\kappa^2 N^2}{(n\omega_0 - \kappa\bar U)^2}
+ \frac{\kappa \bar U''}{n\omega_0 - \kappa\bar U} - \kappa^2 \right] \phi_n = 0,
\label{eq:Ln}
\end{equation}
\noindent with $\phi_n(\kappa,0)=0$, $\phi'_n(\kappa,0)=1$, and $(\cdot)'$
denoting $\dd/\dd z$.  The discrete
spectrum arises from the poles $\phi_n(\kappa,-H)=0$; the eigenvalues
are denoted $k^\pm_{nm}$, $\pm$ denoting rightward/leftward
propagation.  The first-order streamfunction contributed by mode
$(n,m,\pm)$ is
\begin{equation}
\psi^{(1)}_{nm}(x,z,t) = 2\,\mathrm{Re}\!\left\{ A^\pm_{nm}\, \phi_n(k^\pm_{nm},z)\,
\mathrm{exp}({i(k^\pm_{nm}\xi - n\omega_0 t)}) \right\},
\label{eq:psi1_review}
\end{equation}
\noindent with modal amplitude
\begin{equation}
A^\pm_{nm} =  \frac{ \ii  n\omega_0 \, \hat h(k^\pm_{nm}) \,
J_n\!\left(-\dfrac{k^\pm_{nm} U_0}{\omega_0}\right)}
{k^\pm_{nm} \left.\dfrac{\partial\phi_n}{\partial k}\right|_{k^\pm_{nm},\,-H}},
\label{eq:Anm_review}
\end{equation}
\noindent where $n$ is the tidal harmonic, $m$ the modal index, $\hat h$ the
ridge Fourier transform and $J_n$ the Bessel function of the
tide-comoving transformation.  The corresponding velocity and
buoyancy profiles are
\begin{equation}
u^{(1)}_{nm} = 2 \,\mathrm{Re}\left\{ { A^\pm_{nm}\, \phi'_n(k^\pm_{nm},z)\, e^{i\theta_{nm}}} \right\},
\qquad
w^{(1)}_{nm} = -2 \,\mathrm{Re}\left\{ i  k^\pm_{nm}  A^\pm_{nm}\, \phi_n(k^\pm_{nm},z)\, e^{i\theta_{nm}} \right\},
\label{eq:uw}
\end{equation}
\begin{equation}
b^{(1)}_{nm}=2\Real\{A^\pm_{nm}\mathrm B_{nm}(z)\ee^{\ii\theta_{nm}}\},
\qquad
\mathrm B_{nm}(z)=\frac{N^2\phi_{nm}}{\Ue-c_{nm}},
\qquad c_{nm}=\frac{n\omega_0}{k_{nm}},
\label{eq:B1}
\end{equation}
with $\theta_{nm} = k^\pm_{nm}\xi - n\omega_0 t$.

The first-order solution reviewed above is exactly the linear
discrete-spectrum model of \citet{LambDunphy2018}:
equations~(\ref{eq:Ln})--(\ref{eq:B1}) reproduce, with no
modification, their Taylor--Goldstein eigenproblem in the tide-comoving
frame, the Fourier--Bessel decomposition of the topographic forcing and
the residue evaluation of the modal amplitudes.  It constitutes the
baseline onto which the weakly nonlinear corrections of
\S\ref{sec:higher-order} are grafted.  Our numerical implementation of
this first-order solution is validated against the published results of
\citet{LambDunphy2018} in Appendix~\ref{app:lamb}, where their
eigenmode figure is reproduced, and in figure~\ref{fig:ps} below, which
follows the format of their phase-speed figure.

\section{Weakly nonlinear expansion}
\label{sec:higher-order}

Our aim is to integrate the linear model for internal waves foced by bottom ridge with the nonlinear analysis approach. However, it turns out that the previous successive approximation forms \citep{Thorpe1978} are too complicated to allow direct extension to our case. Consequently, we extend the forced architecture of \citet{LambDunphy2018}
(topographic forcing, tide-comoving coordinates, Fourier--Bessel
spectral representation, Taylor--Goldstein resolvent, and residue
evaluation) to $O(\varepsilon^2)$ and $O(\varepsilon^3)$ as follows.  

The strategy still borrows the successive-approximation philosophy of
 finite-amplitude wave analysis --- quadratic
self-interaction at $O(\varepsilon^2)$ and a Fredholm solvability
condition at $O(\varepsilon^3)$ --- but the derivation is reworked from
the governing equations (\ref{eq:me1})--(\ref{eq:master}) for the
present configuration, in which bottom topography and tide-locked
forcing replace freely propagating uniform wavetrain, while reducing the
resulting formulation to  \citet{Thorpe1978}'s free-wavetrain limit provides an
independent check of both the algebra and the numerical implementation. 
These checks are collected in Appendix~\ref{app:thorpe}.  

Unlike  freely propagating uniform wavetrain, the forcing frequency
here is tide-locked; consequently, the $O(\varepsilon^3)$ solvability
condition yields a nonlinear {wavenumber} correction
$\delta k_{nm}$ rather than a frequency correction.  

\subsection{Second-order approximation}\label{sec:second-order}
Following \citet{Thorpe1978}, the second-order problem is obtained by a
systematic weak-nonlinear expansion of the governing vorticity and
buoyancy equations (\ref{eq:me1})--(\ref{eq:me2}).  At order
$\varepsilon^2$, products of two first-order fields appear on the
right-hand side of \eqref{eq:master}, specifically
$-D_t J(\nabla^2\psi_1,\psi_1) + \partial_\xi J(b_1,\psi_1)$.  For each
ordered pair of modes $(p,q)$ with wavenumbers $k_p,k_q$ and harmonic
indices $n_p,n_q$, these products produce a forcing term at the sum
wavenumber $\kappa = k_p+k_q$ and sum frequency
$\nu = \omega_0(n_p+n_q)$.  The nonlinear forcing is decomposed into
two kernels: $K^\nabla_{pq}$ (vorticity-type) and $K^b_{pq}$
(buoyancy-type), given explicitly in
\eqref{eq:knabfix}--\eqref{eq:kbfix}.

Substituting these into the inhomogeneous Taylor--Goldstein equation
yields the second-order forced problem \eqref{eq:S2}, with
inhomogeneous Dirichlet boundary data $g_2$ from the bottom condition.
The self-adjointness of the operator $L_n(\kappa)$ --- the same
property \citet{Thorpe1978} exploited for his orthogonality conditions
--- leads to the Wronskian solvability identity \eqref{eq:wronskian}.
When $\kappa$ coincides with an eigenvalue $k_{nm}$, a bounded solution
exists only if the solvability condition is satisfied; otherwise the
resolvent develops a pole.  The residue of this pole is evaluated via
the Sturm--Liouville identity, yielding the discrete-spectrum amplitude
\eqref{eq:A2} for the second-order wave field.  This procedure mirrors
\citet{Thorpe1978}'s step-by-step construction: knowing the first-order solution,
one computes the quadratic forcing, solves the resulting forced
boundary-value problem, and extracts the second-harmonic and mean-flow
corrections, while enforcing zero-mean displacement to fix the vertical
datum.

\subsubsection{Nonlinear forcing}
\label{sec:kernels}

At order $\varepsilon^2$, the right-hand side of~\eqref{eq:master}
collects products of two first-order fields.  For a general ordered
pair of modes $(p,q)$ with wavenumbers $k_p$, $k_q$ and harmonic
indices $n_p$, $n_q$, define the sum wavenumber $\kappa = k_p + k_q$
and sum frequency $\nu = \omega_0(n_p + n_q)$.  The nonlinear forcing
of (\ref{eq:master}) decomposes as
\begin{equation}
\mathrm{RHS}_2 = \sum_{p,q} 2 \mathrm{Re} \Big\{[i(\nu-\kappa \bar U )K^\nabla_{pq}
+i\kappa K^b_{pq}] e^{i(\kappa\xi-\nu t)}\Big\},
\label{eq:rhs2fix}
\end{equation}
\noindent where the so-called vorticity-type kernel and buoyancy-type
kernel are
\begin{eqnarray}
K^\nabla_{pq}&=& i A_pA_q\Big[k_p\big(\phi_p''-k_p^2\phi_p\big)\phi_q'
-k_q\big(\phi_p'''-k_p^2\phi_p'\big)\phi_q\Big],
\label{eq:knabfix}\\
K^b_{pq}&=&  i A_pA_q\Big[k_p\mathrm B_p\phi_q'-k_q\mathrm B_p'\phi_q\Big],
\qquad \mathrm B_p=\frac{N^2\phi_p}{\bar U-c_p}.
\label{eq:kbfix}
\end{eqnarray}

To derive the above kernels, we start from the second-order forcing,
which for a single ordered mode pair $(p,q)$ takes the form
\[
\mathrm{RHS}_2  = -D_t J(\nabla^2\psi_p,\psi_q) + \partial_\xi J(b_p,\psi_q),
\]
where $\psi_p = A_p \phi_p e^{i\theta_p}$,
$\psi_q = A_q \phi_q e^{i\theta_q}$, and
$b_p = A_p \mathcal{B}_p e^{i\theta_p}$ with
$\mathcal{B}_p = N^2\phi_p/(\bar U - c_p)$ (from the linearized buoyancy
equation).  Evaluating the Jacobians explicitly yields
\[
J(\nabla^2\psi_p,\psi_q) = i A_p A_q \mathcal{R} e^{i(\theta_p+\theta_q)},\quad
J(b_p,\psi_q) = i A_p A_q \mathcal{S} e^{i(\theta_p+\theta_q)},
\]
\noindent with
$\mathcal{R} = k_p(\phi_p''-k_p^2\phi_p)\phi_q' - k_q(\phi_p'''-k_p^2\phi_p')\phi_q$,
$\mathcal{S} = k_p\mathcal{B}_p\phi_q' - k_q\mathcal{B}_p'\phi_q$,
$\theta_p+\theta_q = \kappa\xi-\nu t$, $\kappa=k_p+k_q$, and
$\nu=\omega_0(n_p+n_q)$.  Finally, applying
\[
D_t \rightarrow -i(\nu - \kappa \bar{U}), \qquad \partial_\xi \rightarrow i\kappa
\]
to the operators $-D_t$ and $\partial_\xi$ gives
(\ref{eq:rhs2fix})--(\ref{eq:kbfix}).

The diagonal case $p=q$ reduces to the single-mode self-interaction of
\citet{Thorpe1978}.  Using the first-order Taylor--Goldstein equation
to eliminate the higher $z$-derivatives of $\phi_p$, the diagonal
kernels collapse to the compact forms
\begin{equation}
\mathcal R_{pp}=k\,M'(z)\,\phi^{2},
\qquad
\mathcal S_{pp}=\frac{k\,\big[N^{2}(z)\,\Ue'(z)
-(N^{2})'(z)\,(\Ue-c)\big]}{(\Ue-c)^{2}}\,\phi^{2},
\label{eq:diag}
\end{equation}
where $M(z)$ denotes the coefficient of $\phi$ in the
Taylor--Goldstein operator (\ref{eq:Ln}).  The second term of
$\mathcal S_{pp}$ vanishes for uniform stratification; it is retained
in the code and becomes active in the pycnocline case of
\S\ref{sec:pycnocline}.  Both kernels vanish
identically for uniform flow --- \citet{Thorpe1978}'s uniform-flow limit, which
serves as a built-in consistency check. 

\subsubsection{Inhomogeneous Taylor--Goldstein problem}
\label{sec:inhomTG}
The second-order streamfunction Fourier component $\hat\psi_2(\kappa,z)$
satisfies
\begin{equation}
L_n(\kappa)[\hat\psi_2] = S_2(z;\kappa,n),
\qquad
S_2 = -\frac{i K^\nabla}{\nu - \kappa\bar U}
- \frac{i\kappa K^b}{(\nu - \kappa\bar U)^2},
\label{eq:S2}
\end{equation}
\noindent with boundary conditions
\begin{equation}
\hat\psi_2(\kappa,0) = 0,\qquad
\hat\psi_2(\kappa,-H) = g_2(\kappa,n),
\label{eq:gdata}
\end{equation}
\noindent where $g_2$ follows from the spectral decomposition of the
bottom boundary condition at $O(\varepsilon^2)$, that is,
\begin{equation}
g_2 = i\,\hat w^{\mathrm{BC}}_2(k^\sigma_{nm},n)/k^\sigma_{nm}.
\label{eq:g2}
\end{equation}

The boundary data $g_2$ originates from the bottom condition at
$O(\varepsilon^2)$.  The derivation is summarised below for
completeness.

In the comoving frame, the exact bottom boundary condition is
\[
w = \partial_t \tilde{h} + \left[\bar{U}(z) + u\right] \partial_\xi \tilde{h}
\quad \text{at } z = -H + \tilde{h},
\]
where $\tilde{h} = \varepsilon \bar{h}$ and $\bar{h}=h/\varepsilon$ is
the normalised topography.  Expanding
$w = \varepsilon w_1 + \varepsilon^2 w_2 + \cdots$,
$u = \varepsilon u_1 + \varepsilon^2 u_2 + \cdots$, and Taylor-expanding
about the flat bottom $z=-H$ gives, at first order,
\[
w_1(-H) = \partial_t \bar{h}.
\]
At second order, collecting all $O(\varepsilon^2)$ terms yields
\begin{equation}
\label{eq:w2-1}
w_2(-H) = -\bar{h}\,\partial_z w_1(-H)
+ \left[\bar{U}'(-H)\bar{h} + u_1(-H)\right]\partial_\xi \bar{h}.
\end{equation}

Now substitute the first-order solutions (\ref{eq:uw}):
\[
u_1(\xi,z,t) = 2\, \sum_p \mathrm{Re}\left\{A_p \phi_p'(z) e^{i\theta_p}\right\}, \qquad
w_1(\xi,z,t) = -2\, \sum_p \mathrm{Re}\left\{i k_p A_p \phi_p(z) e^{i\theta_p}\right\},
\]
\noindent where $\theta_p = k_p\xi - n_p\omega_0 t$ and $p$ runs over
all modal indices $(n,m,\pm)$.  Inserting these into (\ref{eq:w2-1})
gives the physical-space expression
\begin{equation}
w_2(-H) = -\bar{h}\,\partial_z w_1(-H)
+ \bar{U}'(-H)\bar{h}\,\partial_\xi \bar{h}
+ u_1(-H)\,\partial_\xi \bar{h}.\label{eq:B3}
\end{equation}

To obtain the spectral form, we apply the Fourier--Bessel
decomposition:
\[
\bar{h}(\xi,t) = \sum_l \hat{h}_l(\xi) e^{-il\omega_0 t}, \qquad
\hat{f}(\kappa) = \int_{-\infty}^{\infty} f(\xi) e^{-i\kappa\xi}\,\mathrm{d}\xi.
\]
The products in (\ref{eq:B3}) become convolutions in wavenumber and
summations over harmonic indices.  Retaining only the component with
sum wavenumber $\kappa$ and sum harmonic $n$ (i.e.\
$\kappa = k_p + k_l$, $n = n_p + n_l$), we define the resulting
spectral coefficient as
\begin{equation}
\hat{w}_2^{\mathrm{BC}}(\kappa,n) \equiv \mathcal{F}\Big\{ -\bar{h}\,\partial_z w_1(-H)
+ \bar{U}'(-H)\bar{h}\,\partial_\xi \bar{h}
+ u_1(-H)\,\partial_\xi \bar{h} \Big\}_{\kappa,n}. \label{eq:B4}
\end{equation}

Finally, using the relation $w = -\psi_\xi$, which in Fourier space
reads $\hat{w}_2 = -i\kappa \hat{\psi}_2$, and evaluating at the
discrete eigenvalue $\kappa = k_{nm}^\sigma$ (where
$\phi_n(k_{nm}^\sigma,-H)=0$), we obtain $g_2$ in (\ref{eq:g2}).

It is also important to distinguish $\mathrm{RHS}_2$ from $S_2$.  The
former is the physical-space forcing that appears on the right-hand
side of the equation (\ref{eq:master}).  The latter is its
spectral-domain counterpart, obtained after Fourier-transforming the
equation for a given frequency--wavenumber pair $(\kappa,\nu)$ and
eliminating the temporal derivatives.  Thus, while $\mathrm{RHS}_2$
contains differential operators acting on products of first-order
fields, $S_2$ is a purely depth-dependent source term for the
inhomogeneous Taylor--Goldstein equation, ready for modal projection
and residue evaluation.

\subsubsection{Solvability identity}
\label{sec:solvability}

The operator $L_n(\kappa)$ with real coefficients is Sturm--Liouville
and self-adjoint on $L^2(-H,0)$ under Dirichlet boundary conditions
--- precisely \citet{Thorpe1978}'s observation for his $L(k,c)$.  Consider
\begin{equation}
L_n[f] = S,\qquad f(0)=0,\quad f(-H)=g,
\label{eq:inhom}
\end{equation}
and let $\Phi(z) \equiv \phi_n(k_{nm},z)$ be the eigenfunction at the
same frequency: $L_n(k_{nm})[\Phi]=0$, $\Phi(0)=\Phi(-H)=0$.
Multiplying~\eqref{eq:inhom} by $\Phi$, the eigenfunction equation by
$f$, and subtracting gives $\Phi S = (\Phi f' - f\Phi')'$.  Integrating
over $[-H,0]$ and using the boundary conditions yields the Wronskian
identity
\begin{equation}
\int_{-H}^0 \Phi(z)\,S(z)\,dz = \Phi'(-H)\,g.
\label{eq:wronskian}
\end{equation}
This is \citet{Thorpe1978}'s orthogonality condition in its general forced form:
when $\kappa$ coincides with an eigenvalue $k_{nm}$ (singular
operator), a bounded solution exists if and only
if~\eqref{eq:wronskian} holds (Fredholm alternative --- the right-hand
side $(S,g)$ must be orthogonal to the nullspace adjoint, which is
$\Phi$ itself).

\subsubsection{Resolvent construction and second-order discrete-spectrum amplitude}
\label{sec:resolvent}

Following Lamb's resolvent construction, the solution of the
inhomogeneous problem~\eqref{eq:inhom} decomposes as
\begin{equation}
f(\kappa,z) = f_p(\kappa,z) + g(\kappa)\,\frac{\phi_n(\kappa,z)}{\phi_n(\kappa,-H)},
\label{eq:decomp}
\end{equation}
where the second term (homogeneous-solution multiple) carries the
inhomogeneous bottom boundary condition (isomorphic to the first-order
$W_n$), and $f_p$ satisfies the non-homogeneous equation with Dirichlet
boundary conditions: $f_p = L_n(\kappa)^{-1}S$.

\textit{(i) Pole structure of the resolvent.} $L_n(\kappa)$ depends
analytically on $\kappa$.  Near an eigenvalue $\kappa = k_{nm}$,
perform the Laurent expansion
$L_n(\kappa) = L_n(k_{nm}) + (\kappa - k_{nm})\partial_k L_n + \cdots$
and set
\begin{equation}
f_p = \frac{a\,\Phi}{\kappa - k_{nm}} + f_{\mathrm{reg}},\qquad
\Phi \equiv \phi_n(k_{nm},\cdot).
\label{eq:laurent}
\end{equation}
Substituting into $L_n(\kappa)[f_p] = S$, the
$O((\kappa-k_{nm})^{-1})$ term vanishes because
$L_n(k_{nm})[\Phi]=0$; the $O(1)$ term gives
$a\,\partial_k L_n[\Phi] + L_n(k_{nm})[f_{\mathrm{reg}}] = S$.
Projecting onto $\Phi$ (self-adjointness makes
$\langle\Phi, L_n(k_{nm})[f_{\mathrm{reg}}]\rangle = 0$) yields the
residue coefficient
\begin{equation}
a = \frac{\langle\Phi, S\rangle}{\langle\Phi, \partial_k L_n\,\Phi\rangle},
\qquad
\langle u,v\rangle \equiv \int_{-H}^0 u(z)\,v(z)\,dz.
\label{eq:residue_a}
\end{equation}

\textit{(ii) Sturm--Liouville identity.} Differentiating
$L_n(k)[\phi_n(k,z)] = 0$ with respect to $k$ gives
$\partial_k L_n[\phi_n] + L_n[\partial_k\phi_n] = 0$.  Projecting the
second term onto $\Phi$ and using the same boundary-integration
technique as for~\eqref{eq:wronskian}:
\begin{equation}
\langle\Phi, \partial_k L_n\,\Phi\rangle
= -\langle\Phi, L_n[\partial_k\phi_n]\rangle
= -\big[ \Phi\,\partial_k\phi'_n - \Phi'\,\partial_k\phi_n \big]_{-H}^0.
\label{eq:sl1}
\end{equation}
At $k = k_{nm}$, $\Phi(0) = \Phi(-H) = 0$.  The normalisation
$\phi_n(k,0)=0$, $\phi'_n(k,0)=1$ holds for all $k$, hence
$\partial_k\phi_n(k,0)=0$.  Therefore
\begin{equation}
\langle\Phi, \partial_k L_n\,\Phi\rangle
= -\Phi'(k_{nm},-H)\,\frac{\partial\phi_n}{\partial k}(k_{nm},-H).
\label{eq:sl2}
\end{equation}
\textit{Verification:} for $\bar U = 0$, $N = \text{const}$, the
left-hand side evaluates to $H/k_{nm}$, and
$-\Phi'(-H)\,\partial_k\phi_n(-H) = H/k_{nm}$ as well --- exact
agreement.  Identity~\eqref{eq:sl2} converts the volume integral
$\langle\Phi,\partial_k L_n\Phi\rangle$ into the product of two bottom
boundary quantities, explaining why only
$\partial\phi_n/\partial k(-H)$ appears in Lamb's residue formula.

\textit{(iii) Residue evaluation.} The two terms in~\eqref{eq:decomp}
each possess a simple pole at $\kappa = k_{nm}$.  Combining them
using~\eqref{eq:sl2}:
\begin{align}
\operatorname*{Res}_{\kappa=k_{nm}} f(\kappa,z)
&= \Phi(z)\left[
\frac{\langle\Phi,S\rangle}{\langle\Phi,\partial_k L_n\Phi\rangle}
+ \frac{g}{\partial\phi_n/\partial k(k_{nm},-H)}
\right] \nonumber\\
&= \Phi(z)\,
\frac{\Phi'(-H)\,g - \langle\Phi,S\rangle}
{\Phi'(-H)\,\partial\phi_n/\partial k(k_{nm},-H)}.
\label{eq:residue_full}
\end{align}
Closing the contour in the upper half-plane for $\xi>0$,
$\int d\kappa\,e^{i\kappa\xi}/(2\pi)$ contributes a factor $i$ (as in
Lamb's first-order derivation).  The second-order discrete-spectrum
field is therefore
\begin{equation}
\psi^{\mathrm{ds}}_2 = 2\, \sum_n \sum_{\pm,m} \mathrm{Re}\left\{
A^{(2)}_{nm}\,\phi_n(k^\pm_{nm},z)\,
e^{i(k^\pm_{nm}\xi - n\omega_0 t)}\right\},
\label{eq:psi2_ds}
\end{equation}
\noindent with modal amplitude
\begin{equation}
A^{(2)}_{nm} = i\,
\frac{\Phi^\pm_{nm}{}'(-H)\,g_2
- \displaystyle\int_{-H}^0 \Phi^\pm_{nm}(z)\,
S_2(z; k^\pm_{nm}, n)\,dz}
{\Phi^\pm_{nm}{}'(-H)\,
\displaystyle\frac{\partial\phi_n}{\partial k}\!\Big|_{k^\pm_{nm},-H}},
\label{eq:A2}
\end{equation}
\noindent where $\Phi^\pm_{nm}(z) \equiv \phi_n(k^\pm_{nm},z)$ and
$g_2 = i\,\hat w^{\mathrm{BC}}_2(k^\sigma_{nm},n)/k^\sigma_{nm}$.

\textit{Diagonal consistency check.} Taking $p = q = (n,m,\sigma)$ for
the sum component ($\kappa = 2k_{nm}$, $\nu = 2n\omega_0$),
substituting~\eqref{eq:knabfix}--\eqref{eq:kbfix}
into~\eqref{eq:S2},~\eqref{eq:A2} and using the first-order
Taylor--Goldstein equation to eliminate the higher $z$-derivatives
recovers the single-mode second-harmonic equation of
\citet{Thorpe1978}.  The general modal-pair formula thus reduces to
 the single-mode limit, simultaneously confirming the
sign conventions in~\eqref{eq:S2}.

\subsubsection{Triad resonance and physical interpretation}
\label{sec:triad}

When the sum wavenumber and frequency of a modal pair fall near an
eigenmode,
\begin{equation}
k_p + k_q \simeq k_{nm},\qquad n_p + n_q = n,
\label{eq:triad}
\end{equation}
\noindent the projection $\langle\Phi_{nm},S_2\rangle$ is resonantly enhanced.
Equation~\eqref{eq:triad} is the discrete-spectrum triad resonance
condition.  Several immediate physical points are summarised as follows. 

\begin{enumerate}
\item \textit{Frequency matching is trivial; wavenumber matching is
non-trivial.}  Tide-locking forces all frequencies to be integer
multiples of $\omega_0$, so the second of~\eqref{eq:triad} is merely
integer addition.  The genuine constraint is the first: the sum of two
parent-mode wavenumbers must equal a free-mode wavenumber --- the
classical three-wave resonance condition $k_1 + k_2 = k_3$,
$\omega_1 + \omega_2 = \omega_3$ in discretised two-dimensional form.

\item \textit{Exact resonance is generally detuned; the response is
dominated by bound harmonics.}  Lamb's eigenvalues are irregularly
distributed under shear and accumulate toward
$n\omega_0/\bar U_{\max}$; exact integer wavenumber matching is
generically absent, with $2k_{nm}$ and $k_p+k_q$ both offset from
eigenvalues.  The inhomogeneous problem~\eqref{eq:S2} then has a
unique, bounded solution; the forcing produces a bound second harmonic
propagating with the parent waves --- \citet{Thorpe1978}'s waveform distortion ---
without resonant growth.  This is one reason Lamb's linear
discrete-spectrum theory remains accurate at finite amplitude.

\item \textit{Near-resonant amplification and detuning length.}  When
$|k_p + k_q - k_{nm}|$ is small but non-zero, the resolvent residue
analysis~\eqref{eq:laurent}--\eqref{eq:residue_a} shows that the local
response scales as $a/(k_p+k_q-k_{nm})$, i.e.\ is amplified.  In the
propagation direction, the near-resonant mode and the forcing beat with
detuning length $\ell_{\mathrm{det}} \sim 2\pi/|k_p+k_q-k_{nm}|$,
causing energy to exchange periodically between parent and harmonic
waves.

\item \textit{The $p=q$ degenerate case is \citet{Thorpe1978}'s self-interaction.}
The diagonal pair yields the $(2k_{nm},2n\omega_0)$ superharmonic
(waveform steepening) and the $p=\bar q$ conjugate pair yields the
$n=0$ mean field (\S\,\ref{sec:meanfield}).

\item \textit{Connection to internal-tide harmonic generation.}
\eqref{eq:triad} is the entry point for harmonic cascades in
internal-tide studies: if $2k_{nm} \simeq k_{2n,m'}$ (second-harmonic
resonance), energy is continuously pumped from the tidal-frequency mode
into the superharmonic mode.  Parametric subharmonic instability (PSI,
$n=1 \to \tfrac12+\tfrac12$) \citep{FanAkylas2021,HilditchThomas2023,AkylasKakoutas2023} does not fall on the integer harmonic grid
in this framework; it belongs to free-beam modulational instability
(requiring a multiple-scale, slowly-varying-amplitude description not
included in the present successive-approximation expansion) and is
listed as future work (\S\,\ref{sec:outlook}).
\end{enumerate}

\subsubsection{Mean field ($n=0$)}
\label{sec:meanfield}

The difference component $p = \bar q$ (conjugate pair) yields $\nu = 0$
forcing.  The operator degenerates to
\begin{equation}
L_0(\kappa) = \frac{d^2}{dz^2} + \frac{N^2}{\bar U^2}
- \frac{\bar U''}{\bar U} - \kappa^2,
\label{eq:L0}
\end{equation}
i.e.\ the long-wave operator for steady lee waves in a background
current \citep{Bell1975}.  Two points merit attention:
\begin{enumerate}
\item $L_0$ is singular where $\bar U \to 0$ (near the bottom of Lamb's
profile), so the mean-field response involves critical-layer-type
structures near $\bar U = 0$, directly connected to the
continuous-spectrum problem \citep{OnukiVenaille2025}.
\item \citet{Thorpe1978} eliminated the $O(\varepsilon^2)$ mean flow in a free
wavetrain via a ``zero mean displacement'' condition.  In the forced
topographic problem, the second-order mean field is a genuine physical
field --- the steady lee-wave response and the source of the mean
energy-flux corrections.  Its diagonal (self-interaction) component is
computed here through the Long-equation expansion: $e_{20}$ of
(\ref{eq:e20}) is the mean displacement field, which enters the
third-order solvability (\ref{eq:c2}).  The singular $\bar U\to0$
critical-layer structure of the general $n=0$ problem and the
off-diagonal (mode-pair) mean fields are left to future work
(\S\,\ref{sec:outlook}).
\end{enumerate}

\subsection{Third-order approximation}
\label{sec:third-order}

At $O(\varepsilon^3)$:
\begin{align}
M\psi_3 &= -D_t\big[ J(\nabla^2\psi_1,\psi_2) + J(\nabla^2\psi_2,\psi_1) \big]
+ \partial_\xi\big[ J(b_1,\psi_2) + J(b_2,\psi_1) \big], \label{eq:M3}\\
D_t b_3 &= N^2 \psi_{3\xi} - J(b_1,\psi_2) - J(b_2,\psi_1),
\label{eq:b3}
\end{align}
with the same boundary-condition structure as~\eqref{eq:gdata}.  Under
the fundamental-frequency truncation, the forcing contains $n=1$
components ($E_1\times E_0$ and $E_2\times E_{-1}$) and $n=3$
components ($E_1\times E_2$).

\subsubsection{Resonant component and nonlinear wavenumber correction}
\label{sec:delta_k}

The $n=1$ forcing at $\kappa = k_{nm}$ is resonant with the homogeneous
operator: generically~\eqref{eq:wronskian} is not satisfied, the forced
problem has no bounded solution, and secular terms $\propto \xi$
appear --- unacceptable in the radiation zone.  The treatment for
a free wavetrain allows the frequency (phase speed) to acquire an
$O(\varepsilon^2)$ correction.  In the Lamb forced problem, the
frequency is tide-locked to $n\omega_0$; the corresponding degree of
freedom is the modal wavenumber:
\begin{equation}
k_{nm} \;\longrightarrow\; \tilde k_{nm} = k_{nm} + \varepsilon^2\delta k_{nm},
\qquad
\tilde\theta_{nm} = \tilde k_{nm}\xi - n\omega_0 t.
\label{eq:ktilde}
\end{equation}

\textit{Derivation.}  With the corrected wavenumber, the first-order
mode $\varepsilon A_{nm}\phi_n(k_{nm},z)e^{i\tilde\theta}$ is no longer
on-shell.  Acting with $M$ from~\eqref{eq:master} gives
$-\varepsilon(n\omega_0 - \tilde k_{nm}\bar U)^2 L_n(\tilde k_{nm})[\phi_n(k_{nm})]e^{i\tilde\theta}$.
Expanding
$L_n(\tilde k_{nm}) = L_n(k_{nm}) + \varepsilon^2\delta k_{nm}\,\partial_k L_n + \cdots$,
the first-order field leaves an $O(\varepsilon^3)$ residual
\begin{equation}
-\varepsilon^3\delta k_{nm}\,(n\omega_0 - k_{nm}\bar U)^2\,
A_{nm}\,\partial_k L_n(k_{nm})[\Phi_{nm}]\,e^{i\theta_{nm}}.
\label{eq:residual}
\end{equation}

Moving this into the right-hand side of the $n=1$ resonant component
of~\eqref{eq:M3} (equivalent to adding
$-\delta k_{nm}A_{nm}\partial_k L_n[\Phi_{nm}]$ to $S_3$ after division
by $-(\nu-\kappa\bar U)^2$) and imposing the solvability
condition~\eqref{eq:wronskian}:
\begin{equation}
\delta k_{nm}\,A_{nm}\!\int_{-H}^0 \Phi_{nm}\,\partial_k L_n[\Phi_{nm}]\,dz
= \int_{-H}^0 \Phi_{nm}\,S_3^{\mathrm{res}}(z)\,dz
- \Phi'_{nm}(-H)\,\hat g_3^{\mathrm{res}}.
\label{eq:dk_solv}
\end{equation}

Using identity~\eqref{eq:sl2} to convert the denominator to boundary
quantities yields the compact form
\begin{equation}
\delta k_{nm} = -\,\frac{\displaystyle\int_{-H}^0 \Phi_{nm}(z)\,S_3^{\mathrm{res}}(z)\,dz
- \Phi'_{nm}(-H)\,\hat g_3^{\mathrm{res}}}
{A_{nm}\,\Phi'_{nm}(-H)\,
\displaystyle\frac{\partial\phi_n}{\partial k}(k_{nm},-H)},
\label{eq:delta_k_final}
\end{equation}
\noindent where $S_3^{\mathrm{res}}$ is the projection of the right-hand side
of~\eqref{eq:M3} onto $(k_{nm},n\omega_0)$ (composed of cross products
of the first-order modes with their second-order corrections), and
$\hat g_3^{\mathrm{res}}$ is the corresponding component of the
$O(\varepsilon^3)$ boundary data.  Explicitly,
\begin{equation}
\partial_k L_n(k) = \frac{2kN^2}{(n\omega_0 - k\bar U)^2}
+ \frac{2k^2 N^2 \bar U}{(n\omega_0 - k\bar U)^3}
+ \frac{\bar U''}{n\omega_0 - k\bar U}
+ \frac{k\bar U''\bar U}{(n\omega_0 - k\bar U)^2} - 2k.
\label{eq:dkLn}
\end{equation}

The corresponding phase-speed correction (Stokes-type nonlinear
dispersion) is
\begin{equation}
\delta c_{nm} = -\frac{n\omega_0}{k_{nm}^2}\,\delta k_{nm},
\qquad
c_{nm}^{\mathrm{eff}} = c_{nm} + \varepsilon^2\delta c_{nm},
\label{eq:delta_c}
\end{equation}
\noindent which corresponds term-by-term to \citet{Thorpe1978}'s finite-amplitude
phase-speed shift; the only difference is that the correction manifests
in wavenumber rather than frequency.

\textit{Physical consequences.}  (1)~Modal wavelength and phase speed
acquire amplitude-dependent corrections: at fixed tidal frequency
$|k|$ increases with amplitude, so the effective phase speed
$|c^{\rm eff}|=n\omega_0/|k+\varepsilon^2\delta k|$ decreases.
(2)~$\delta k_{nm} \propto |A_{nm}|^2$ (since $S_3^{\mathrm{res}}$
contains $|A|^2 A$-type cubic terms), consistent with \citet{Thorpe1978}'s $a^2$
dependence.

\subsubsection{Non-resonant component: third harmonic}
\label{sec:third-harmonic}

The $n=3$ forcing ($E_1 \times E_2$ type) at $(3k_{nm},3n\omega_0)$
yields the inhomogeneous boundary-value problem
\begin{equation}
L_{3n}(3k_{nm})[\Psi_{3,nm}] = S_3^{(E_3)}(z),\qquad
\Psi_{3,nm}(0) = \Psi_{3,nm}(-H) = 0.
\label{eq:Y3}
\end{equation}
In general $3k_{nm}$ is \emph{not} an eigenvalue of $3n\omega_0$, so
the solution is unique --- the forced counterpart of \citet{Thorpe1978}'s
third-harmonic correction.  A noteworthy internal-resonance possibility
exists: if a mode's third harmonic happens to satisfy
$3k_{nm} \approx k_{3n,m'}$ (a special case of the triad
condition~\eqref{eq:triad}, known in internal-tide studies as
harmonic resonance), the response is resonantly amplified, again
requiring the solvability condition~\eqref{eq:wronskian}.  This
provides the entry point for studying harmonic cascades within the Lamb
architecture.

Assembling first through third order, the single-mode wave profile is
\begin{equation}
\psi \simeq 2\,\Real\!\Big\{
\tilde A_{nm}\Phi_{nm}\,e^{i\tilde\theta_{nm}}
+ \tilde A_{nm}^2\Psi_{2,nm}\,e^{2i\theta_{nm}}
+ \tilde A_{nm}^3\Psi_{3,nm}\,e^{3i\theta_{nm}}
\Big\},
\label{eq:psi123}
\end{equation}
with $\tilde A_{nm} = \varepsilon A_{nm}$.  In the multi-mode case,
sums over all mode pairs are taken and the second-order mean field
$\psi_2^{(n=0)}$ (\S\,\ref{sec:meanfield}) is added.

\subsection{Single-mode diagonal reduction}
\label{sec:diagonal}

The computations of \S\,\ref{sec:results} retain the diagonal (self-interaction) part of the general expansion only. For a single mode $(n,m,\pm)$ with complex amplitude $A$, the
diagonal kernels~\eqref{eq:diag} give the source
\begin{equation}
S_2 = A^{2}\left[\frac{\mathcal R_{pp}}{\nu-\kappa\Ue}
+ \frac{\kappa\,\mathcal S_{pp}}{(\nu-\kappa\Ue)^{2}}\right],
\qquad \kappa=2k,\quad \nu=2n\omega_0,
\label{eq:S2diag}
\end{equation}
\noindent and the bound second harmonic $\hat\Psi_2$ solves the forced Taylor--Goldstein problem
\begin{equation}
L_{2n}(2k)[\hat\Psi_2]=S_2,
\qquad
\hat\Psi_2(0)=\hat\Psi_2(-H)=0,
\label{eq:Psi2}  
\end{equation}
with the associated buoyancy profile
\begin{equation}
\hat{\mathrm B}_2=\frac{A^2\mathcal S_{pp}-\kappa N^2\hat\Psi_2}{\nu-\kappa\Ue},
\qquad
b_2=2\Real\{\hat{\mathrm B}_2\ee^{2\ii\theta}\}.
\label{eq:B2}
\end{equation}
The $O(\varepsilon^2)$ inhomogeneous bottom-boundary forcing (the $g_2$
term of the general theory) is neglected here: it drives an
independent, Lamb-type linear response at $2n\omega_0$ rather than the
bound waveform distortion, and is left to future work
(\S\,\ref{sec:outlook}).

At third order the resonant source contains, besides the cross
interaction of the fundamental with its bound second harmonic, the
interaction with the second-order \emph{mean field} the mode generates
in itself.  Assembling these cross Jacobians directly in the
$(\psi,b)$ variables proved error-prone and, more importantly, such an
assembly omits the mean-field contribution altogether.  We therefore
compute the resonant projection from the exact Long equation for the
isopycnal vertical displacement
$\eta(\xi,z)$ in a steady parallel current,
\begin{equation}
\nabla^2\eta + \mathcal B(w)\,\eta
- \mathcal D(w)\big(\eta_\xi^2 - 2\eta_z + \eta_z^2\big) = 0,
\qquad w = z-\eta,
\label{eq:long}
\end{equation}
\noindent with $\mathcal B(z)=N^2/(\Ue-c)^2$ and
$\mathcal D(z)=\Ue'/(\Ue-c)$.  Expanding
$\eta=\varepsilon\eta_1+\varepsilon^2\eta_2+\cdots$ about a single
mode, the first-order displacement is written
$\eta_1=H(z)\cos k\xi$ with the displacement eigenfunction
\begin{equation}
H(z)=\frac{\phi(z)}{c-\Ue(z)},
\qquad \max_z|H|=1,
\label{eq:H}
\end{equation}
\noindent so that $\varepsilon=\max_{z,\xi}|\eta_1|$ is the
displacement amplitude.  At $O(\varepsilon^2)$ the mean field
$e_{20}(z)$ and the bound second harmonic $e_{22}(z)$, defined by
$\eta_2=e_{20}(z)+e_{22}(z)\cos2k\xi$, solve
\begin{equation}
e_{20}''+2\mathcal D e_{20}'+\mathcal B e_{20}=S_0,
\qquad
S_0=\frac{\mathcal B'H^2}{2}
+\frac{\mathcal D\,(k^2H^2+H'^2)}{2}+\mathcal D'HH',
\label{eq:e20}
\end{equation}
\begin{equation}
e_{22}''+2\mathcal D e_{22}'+(\mathcal B-4k^2)e_{22}=S_2,
\qquad
S_2=\frac{\mathcal B'H^2}{2}
+\frac{\mathcal D\,(H'^2-k^2H^2)}{2}+\mathcal D'HH',
\label{eq:e22}
\end{equation}
\noindent with $e_{20}=e_{22}=0$ at $z=0$ and $z=-H$.  The
second-harmonic problem (\ref{eq:e22}) is equivalent to the forced
Taylor--Goldstein solution (\ref{eq:Psi2})--(\ref{eq:B2}); the mean
field $e_{20}$ is the component that a cross-Jacobian assembly in the
$(\psi,b)$ variables misses.  For constant $N$ and $\Ue\equiv0$ both
sources vanish identically, an analytic regression check of the
implementation.

At $O(\varepsilon^3)$, solvability of the resonant fundamental against
the adjoint weight $\mu=s^2H$, with $s=\Ue-c$, fixes the phase-speed
correction $c=c_0+\varepsilon^2 c_2$ as
\begin{equation}
c_2=-\frac{N_3}{D_3},
\qquad
D_3=\int_{-H}^{0}\frac{2N^2}{s}\,H^2\,\dd z,
\label{eq:c2}
\end{equation}
\begin{align}
N_3=\int_{-H}^{0}s^2H\Big\{&
-2\mathcal B'H\big(e_{20}+\tfrac12 e_{22}\big)
+\tfrac{3}{8}\mathcal B''H^3
-2\mathcal D\big(k^2He_{22}+H'e_{20}'+\tfrac12 H'e_{22}'\big)
\notag\\
&+\mathcal D'\Big[\tfrac12 H\big(k^2H^2+H'^2\big)
+\tfrac14 H\big(H'^2-k^2H^2\big)
-2He_{20}'-He_{22}'\Big]
\notag\\
&+\tfrac34\mathcal D''H^2H'
-2\mathcal D'H'e_{20}
-\mathcal D'H'e_{22}
\Big\}\,\dd z .
\label{eq:N3}
\end{align}

For the forced, tide-locked problem the frequency is fixed at
$n\omega_0$, so the correction migrates from the phase speed to the
wavenumber: expanding the modal-branch dispersion
$\omega(k,\varepsilon)=k\,[c_0+\varepsilon^2 c_2]$ at fixed
$\omega=n\omega_0$ gives
\begin{equation}
\delta k_{nm}=-\frac{k_{nm}}{c_g}\,\varepsilon^2 c_2,
\qquad
\varepsilon\equiv\max_{z,\xi}|\eta_1|
=2|A_{nm}|\max_z\left|\frac{\phi}{c-\Ue}\right|,
\label{eq:dklong}
\end{equation}
\noindent where the factor of two reflects the $2\Real$ convention of
(\ref{eq:psi1_review}) and $c_g=\dd\omega/\dd k$ is the group velocity
along the modal branch, obtained by finite-differencing the shooting
dispersion function of \S\ref{sec:first-order-review}.  This is the
explicit realisation of the general form (\ref{eq:delta_k_final}):
since $c_2$ is a real functional of the mode structure,
$\delta k_{nm}\propto|A_{nm}|^2$ and is purely real.

For all profiles of this paper the projection gives $c_2<0$, and
$\delta k$ then carries the sign of $k$: $|k|$ grows with amplitude
and the effective phase speed $|c^{\rm eff}|=n\omega_0/|k+\delta k|$
\emph{decreases} --- a finite-amplitude slow-down, in agreement with
the sign of \citet{Thorpe1978}'s central-branch correction and opposite to the
sign produced by the earlier cross-Jacobian assembly.


\section{Results \& discussion}
\label{sec:results}

In the remainder of this work, we first replicate the results of \citet{LambDunphy2018} using identical set-ups, which serves as a validation of our numerical inplementation of the theoretical model. We then examine the nonlinear effects within these configurations; the negligible discrepancies observed demonstrate why the linear model (with the first-order solution) works so effectively. Finally, we turn to the pycnocline case to further quantify these nonlinear effects, still confining ourselves to the weak internal waves considered herein.

\subsection{The set-up with uniform $N$}
\begin{table}
  \begin{center}
    \begin{tabular}{lll}
      Parameter & Value & Meaning \\
      \midrule
      $H$ & $5000\,$m & water depth ($z=0$ rigid lid, $z=-H$ flat bottom) \\
      $N$ & $10^{-3}\,$rad\,s$^{-1}$ & constant buoyancy frequency \\
      ${\omega_0}$ & $2\pi/(12.42\times3600)$ & M$_2$ tidal frequency $\omega_0$ \\
      $U_0$ & $0.05\,$m s$^{-1}$ & barotropic tidal current amplitude $U_0$ \\
      ${h_0}$ & $0.1\,H$ & ridge height (small parameter $\varepsilon=h_0/H=0.1$) \\
      ${a}$ & $1.2\,H$ & Gaussian ridge half-width \\
      ${U_s}$ & $0.5\,$m s$^{-1}$ (set 2), $0.16\,$m s$^{-1}$ (set 7) & surface current $U_s$ \\
      ${z_s}$ & $-0.3\,H$ & shear-layer centre $z_s$ \\
      ${d_s}$ & $0.06\,H$ & shear-layer thickness $d_s$ \\
    \end{tabular}
    \caption{Parameters used in the baseline simulations, following
    \citet{LambDunphy2018}.}
    \label{tab:params}
  \end{center}
\end{table}

{The linear ingredients (eigenvalue scans, eigenfunctions, and
$\partial_k\phi_n/\partial k$ at the bottom) are computed with our own
validated discrete-spectrum code. } {We use the standard parameters of \citet{LambDunphy2018}: $H=5000\,$m,
$N=10^{-3}$~rad\,s$^{-1}$, M$_2$ tide $\omega_0=2\pi/(12.42\,{\rm h})$,
$U_0=0.071H\omega_0$, $h_0/H=0.1$, Gaussian ridge $a/H=1.2$; two
background currents are considered: sets~1--3
($U_s/(NH)$, $z_s/H$, $d_s/H$) $=(0.1,-0.3,0.06)$ and set~7
$=(0.16,-0.3,0.06)$.  At least ten modes per propagation direction are retained
for harmonic $n=1$.  


{As mentioned above, figure~\ref{f:fig1}  is about the effect of the $n=2$ harmonic. We first assess the contribution of the second harmonic to the surface flow field. For the surface current, the maximum amplitude of the fundamental mode alone is $\max |u_{(n=1)}| = 4.97 \times 10^{-3}\, (NH)$, while including the $n=2$ component yields $\max |u_{(n=1+2)}| = 4.98 \times 10^{-3}\, (NH)$. The difference between the two corresponds to approximately $3.8\%$ of the fundamental amplitude. For the present set-up, the fundamental linear solution therefore overwhelmingly dominates the flow field. 

Figure~\ref{fig:ps} further shows the normalised phase speeds $|c|/(NH)$ of the first ten
vertical modes for the two background currents, with the
third-order corrected values $c^{\rm eff}=n\omega_0/(k+\delta k)$
overlaid as open squares.  The corrections are invisible at the scale
of the plot.  This is the first
quantitative answer to our question --- at the parameters of
\citet{LambDunphy2018} the nonlinear dispersion correction is far below
the accuracy at which the linear theory has been tested against
simulations.
\begin{figure}
\centering
\includegraphics[width=0.72\textwidth]{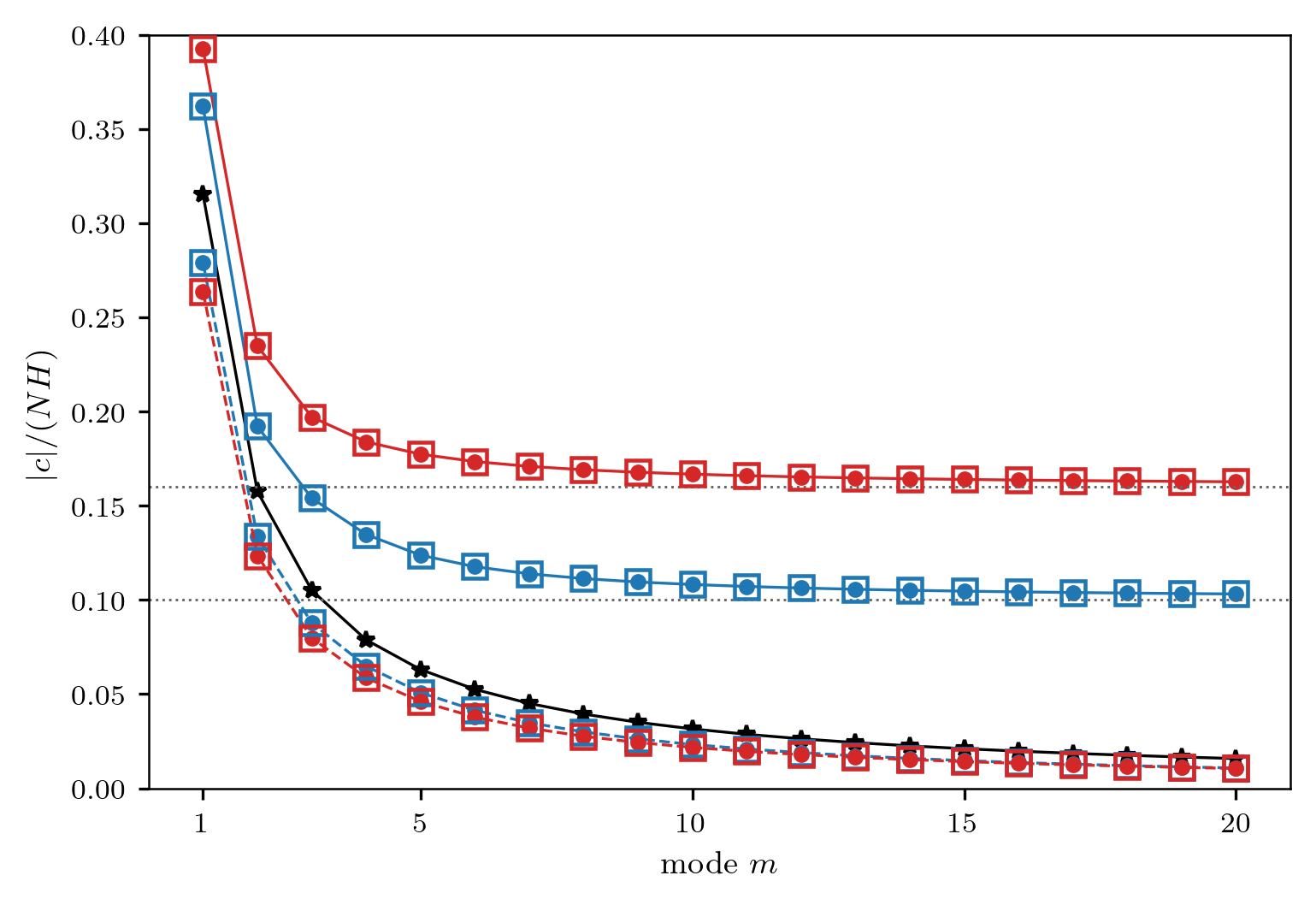}
\caption{Phase speeds $|c|/(NH)$ versus vertical mode number for the constant $N$ case: (solid curves) rightward propagating waves, (dashed curves) leftward waves, for the sets~1--3 ($U_s/(NH)=0.1$, blue) and set~7 ($0.16$, red) currents; (black stars)
no current; (dotted lines) the limiting values $\Ue^{\max}/(NH)=0.1,
0.16$.  (Filled circles) linear theory; (open squares) with the
third-order nonlinear wavenumber correction $c^{\rm eff}$ of (\ref{eq:delta_c}), indistinguishable at this scale (the corrections
are further quantified in figure~\ref{fig:dk}).}
\label{fig:ps}
\end{figure}

The corrections themselves are shown in figure~\ref{fig:dk} separately. They peak at
modes 2--3, reaching $|\delta k|\,H=4.43\times10^{-5}$ (sets~1--3, 
rightward mode~2) and $1.17\times10^{-4}$ (set~7, leftward mode~2),
i.e.\ $|\delta c|/(NH)=c\,|\delta k|/|k|$ up to $1.27\times10^{-5}$; the correction is
largest for the stronger current, growing by a factor of $\simeq2.6$
between sets~1--3 and set~7.  {While the corrections are numerically small, two structural features are nonetheless discernible.} First,
$\delta k/k$ is positive for essentially all modes of both directions: the nonlinear correction slightly increases $|k|$ and hence reduces $|c|$ at fixed tidal frequency --- a uniform nonlinear slow-down.  Second, the modal decay differs sharply by direction: for leftward modes the
correction collapses beyond mode~5, tracking the modal amplitude $|A_{nm}|$ which is cut off by the ridge spectrum
$\hat h\propto\ee^{-(ak)^2/4}$ at large $|k|$, whereas for rightward modes, whose wavenumbers accumulate at $n\omega_0/\Ue^{\max}$ and whose amplitudes decay only slowly, the correction persists at the
$|\delta k|\,H\sim10^{-6}$ level out to mode~10.  Since $\delta k\propto|A|^2\propto(h_0/H)^2$, these numbers scale
quadratically with ridge height.
\begin{figure}
\centering
\includegraphics[width=0.72\textwidth]{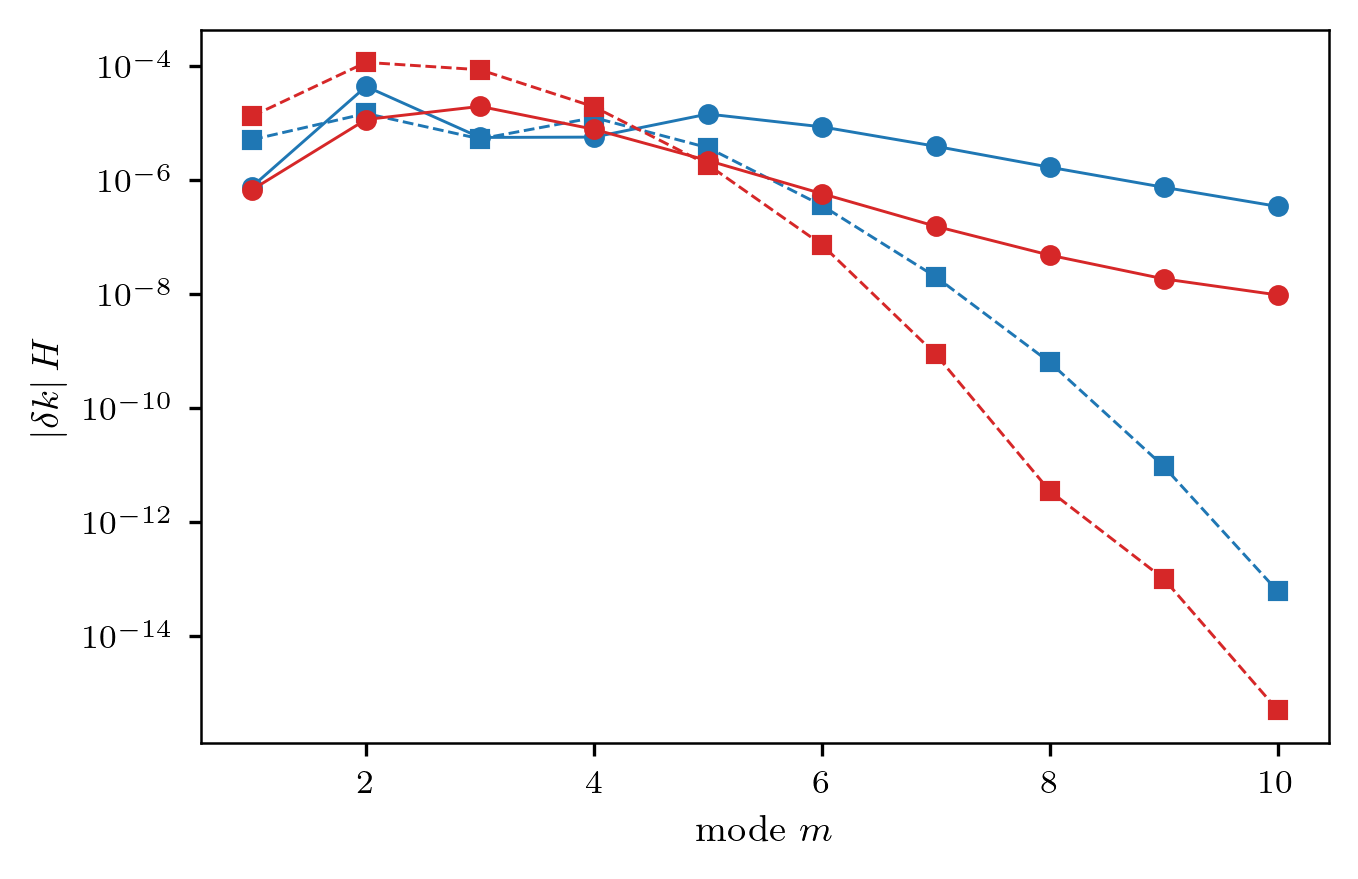}
\caption{Wavenumber correction $|\delta k|\,H$ versus mode number in the
$H$-normalised form of figure~6 of \citet{LambDunphy2018}.  Blue:
sets~1--3 current; red: set~7; circles/solid: rightward modes;
squares/dashed: leftward modes.  The corrections peak at modes 2--3,
grow with current strength, and decay with mode number far more slowly
for rightward than for leftward modes (see text).}
\label{fig:dk}
\end{figure}

Next, figure~\ref{fig:wave} shows the finite-amplitude waveform in the current
setting with constant $N$: the vertical profile of
the isopycnal vertical displacement $\eta=b/N^2$ at the crest phase for modes 1,
2, 4 and 8 of the sets~1--3 current, both propagation directions.  The
profiles are normalised by $\max|\eta_1|$ for three reasons: the
absolute amplitudes differ by up to two orders of magnitude between
modes (see the caption), so no common dimensional scale would show all
waveforms; the linear profile scales linearly with $h_0$, so its
normalised shape is amplitude-independent; and the relative distortion
$\eta_2/\eta_1$ is the intrinsic nonlinear measure.  The bound second
harmonic $\eta_2$ (dashed) reaches 0.3\%--2.3\% of the linear
displacement $\eta_1$ (up to 2.9\% for set~7 rightward mode~4).  The
red total is therefore barely distinguishable from the linear profile
--- consistent with the phase-speed result --- but the {structure}
of $\eta_2$ is physically informative: for rightward modes the
distortion peaks at $z/H\simeq-0.15$, inside the surface shear layer
(centred at $z_s/H=-0.3$), where both $\Ue'$ and the eigenfunction are
large; for leftward modes it peaks much deeper ($z/H\simeq-0.8$),
following the modal displacement.  The shear layer is thus confirmed as
the seat of the finite-amplitude distortion of the
downstream-propagating tide.  The
mode-8 panels add a further directional contrast: the leftward mode~8
is essentially unforced ($\max|\eta_1|=0.03$~m, cut off by the ridge
spectrum at large $|k|$), whereas the rightward mode~8, approaching the
spectral accumulation point $c\to\Ue^{\max}$, still carries
$\max|\eta_1|\simeq1$~m with a distortion ratio of 1.3\% --- larger
than for modes~1 and~4.

\begin{figure}
\centering
\includegraphics[width=0.98\textwidth]{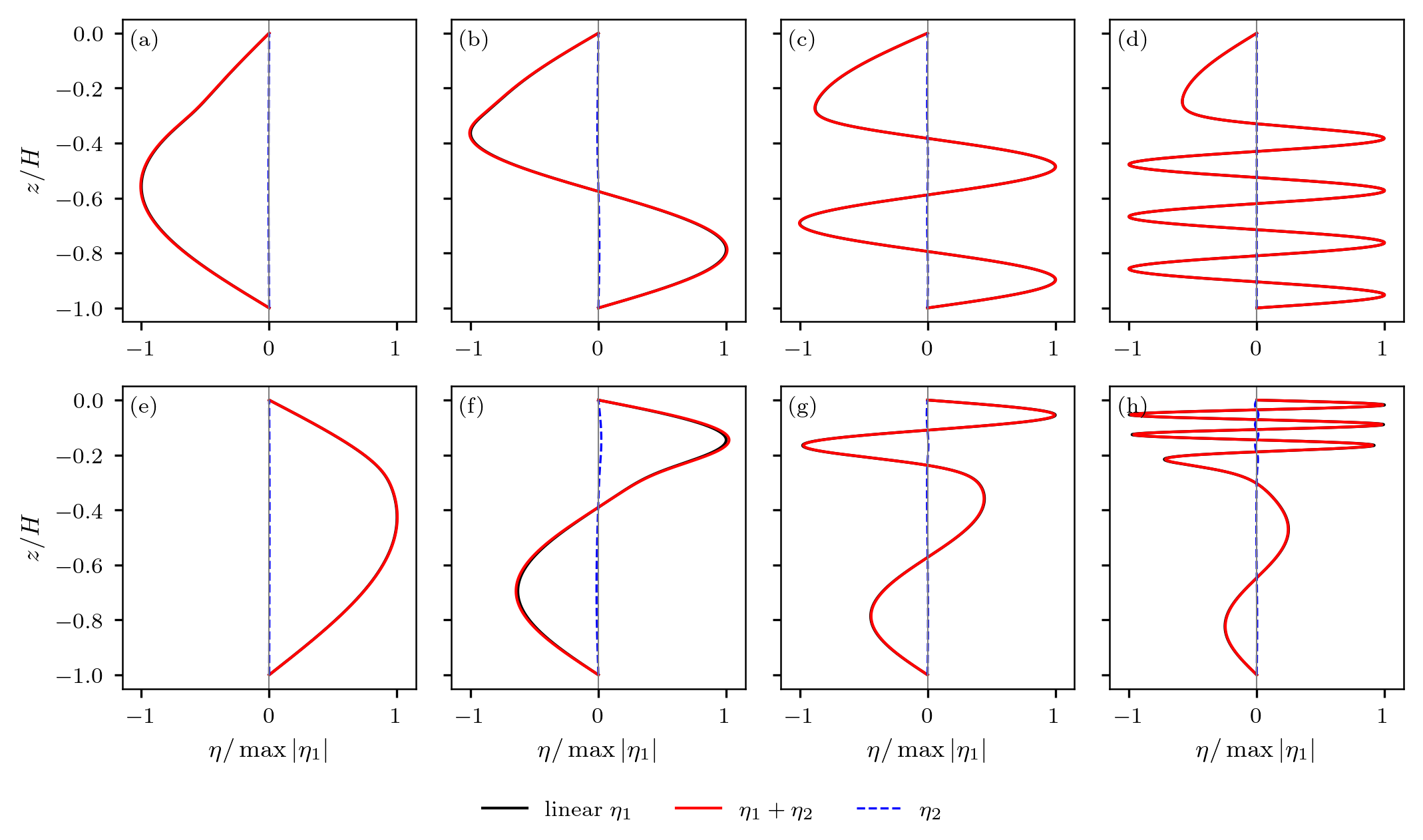}
\caption{Displacement profiles $\eta(z)=b/N^2$ at the crest phase,
normalised by $\max|\eta_1|$ (see text), for the sets~1--3 current.
(a)--(d) leftward modes $m=1,2,4,8$; (e)--(h) rightward modes
$m=1,2,4,8$.  Black: linear first-order profile $\eta_1$; blue dashed:
bound second harmonic $\eta_2$; red: total.  Per-mode data ($c$ in
m\,s$^{-1}$; $\delta k/k$; distortion ratio
$\max|\eta_2|/\max|\eta_1|$; $\max|\eta_1|$ in m): leftward $m=1$:
$-1.40$, $+1.0\times10^{-5}$, $5.5\times10^{-3}$, $5.97$; $m=2$:
$-0.67$, $+1.4\times10^{-5}$, $7.9\times10^{-3}$, $9.31$; $m=4$:
$-0.32$, $+5.8\times10^{-6}$, $3.4\times10^{-3}$, $5.57$; $m=8$:
$-0.15$, $\simeq0$, $2.9\times10^{-5}$, $0.03$; rightward $m=1$:
$+1.81$, $+2.0\times10^{-6}$, $2.7\times10^{-3}$, $3.25$; $m=2$:
$+0.96$, $+6.1\times10^{-5}$, $2.3\times10^{-2}$, $5.17$; $m=4$:
$+0.67$, $+5.5\times10^{-6}$, $6.5\times10^{-3}$, $3.32$; $m=8$:
$+0.56$, $+1.3\times10^{-6}$, $1.3\times10^{-2}$, $0.99$.}
\label{fig:wave}
\end{figure}

\subsection{Stratification co-located with the shear layer}
\label{sec:pycnocline}

All results above use the uniform stratification (with constant buoyancy frequency $N=10^{-3}$) of
\citet{LambDunphy2018}.  \citet{Thorpe1978}'s own wave breaking study, in contrast,
took the density and velocity transitions to be co-located --- his
basic state $\Ue=U_1\tanh p(z-d)$, $\rho=\rho_0[1-A\tanh p(z-d)]$
puts a strong buoyancy-frequency maximum inside the shear layer, as in
the oceanic pycnocline.  Below, we examine this configuration to isolate the effect of
 $N(z)$, while keeping all other
parameters identical to those in the constant-$N$ case,
\begin{equation}
N^2(z) = N_b^2 + B^2\,\mathrm{sech}^2\!\big((z-z_s)/d_s\big),
\qquad
N_b^2 = 0.1\,N_{\rm ref}^2,
\label{eq:pycnocline}
\end{equation}
\noindent co-located with the shear layer ($z_s/H=-0.3$, $d_s/H=0.06$), with
$B^2=7.5\,N_{\rm ref}^2$ chosen so that the depth-mean $N^2$ still equals
that of the constant-$N$ reference, i.e., $N_{\rm ref}=10^{-3}$~rad/s 
(see figure~\ref{fig:pyc_setup}).  The peak stratification is
$7.6\,N_{\rm ref}^2$ ($N_{\max}\simeq2.8\,N_{\rm ref}$); because the
buoyancy and velocity gradients now coincide, the minimum background
Richardson number {rises} from $1.03$ (the former constant $N$ cases) to $6.5$ ---
the pycnocline shear layer is further from Kelvin--Helmholtz
instability even as it becomes the preferred seat of wave activity.
%
%
\begin{figure}
\centering
\includegraphics[width=0.98\textwidth]{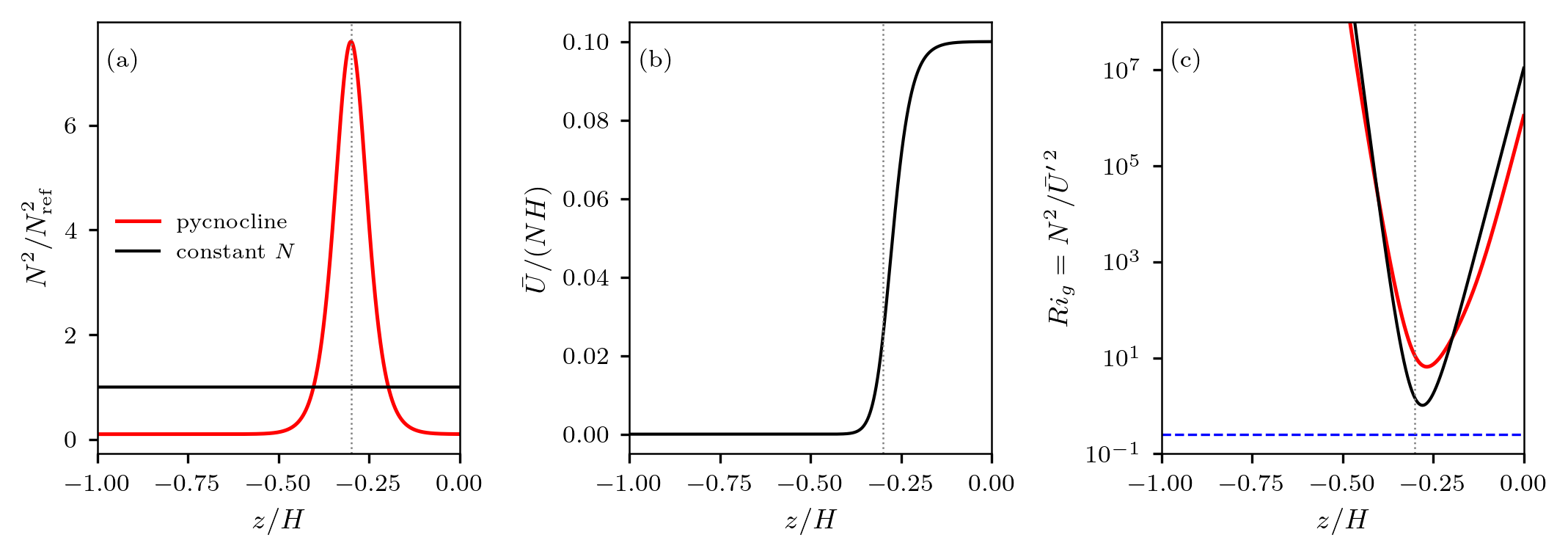}
\caption{The new  pycnocline case (\ref{eq:pycnocline}):
(a) $N^2(z)/N_{\rm ref}^2$ (red) versus the constant-$N$ reference
(black); (b) the sets~1--3 background current (common to both cases);
(c) background gradient Richardson number $Ri_g=N^2(z)/\Ue'^{\,2}$.
Dotted line: shear-layer centre $z_s/H=-0.3$; dashed blue:
Miles--Howard criterion $1/4$.  In contrast to the previous case set-ups from \citet{LambDunphy2018}, 
the new case considered here co-locates the buoyancy and velocity gradients, which raises the minimum Richardson number $\min Ri_g$ from $1.03$ to $6.5$.}
\label{fig:pyc_setup}
\end{figure}
The solution procedure is as follows. We first solve the homogeneous Taylor--Goldstein equation (\ref{eq:Ln}) to
obtain the discrete eigenvalues $k^\pm_{nm}$. With these eigenvalues, the first-order linear modal amplitude is computed from (\ref{eq:Anm_review}). The resulting first-order field then provides the forcing term $S_2$ in (\ref{eq:S2diag}), which drives the inhomogeneous Taylor--Goldstein equation  (\ref{eq:Psi2}). Superposing the solutions of the homogeneous and inhomogeneous problems yields the full second-order wave field. It should be noted that the physics targeted here is precisely about the diagonal part, because the waveform-distortion analysis \citep{Thorpe1978} 
is for the degenerate $p=q$ case, where self-interaction produces the                                                                        
     $(2k_{nm},2n\omega_0)$ superharmonic (waveform steepening), and the                                                         
     conjugate pair $p=\bar q$ yields the $n=0$ mean field.  

  Off-diagonal contributions are higher-order in amplitude.                                              
     The amplitude of an off-diagonal bound wave scales as                                                                       
     $|A_p||A_q|$, whereas the modal spectrum is strongly concentrated in                                                        
     the lowest modes (for the current ridge set-up with $a/H=1.2$, modes $m=2,3,1$ dominate).  Since                                                            
     $|A_pA_q|\leq\tfrac12(|A_p|^2+|A_q|^2)$, even the largest off-diagonal                                                      
     term cannot exceed the diagonal terms of the dominant modes, and the                                                        
     vast majority of off-diagonal pairs are far smaller.


\begin{figure}
\begin{center}
\begin{minipage}{130mm}
\begin{center}
    \subfigure[ ]
              {\includegraphics[width=120mm]{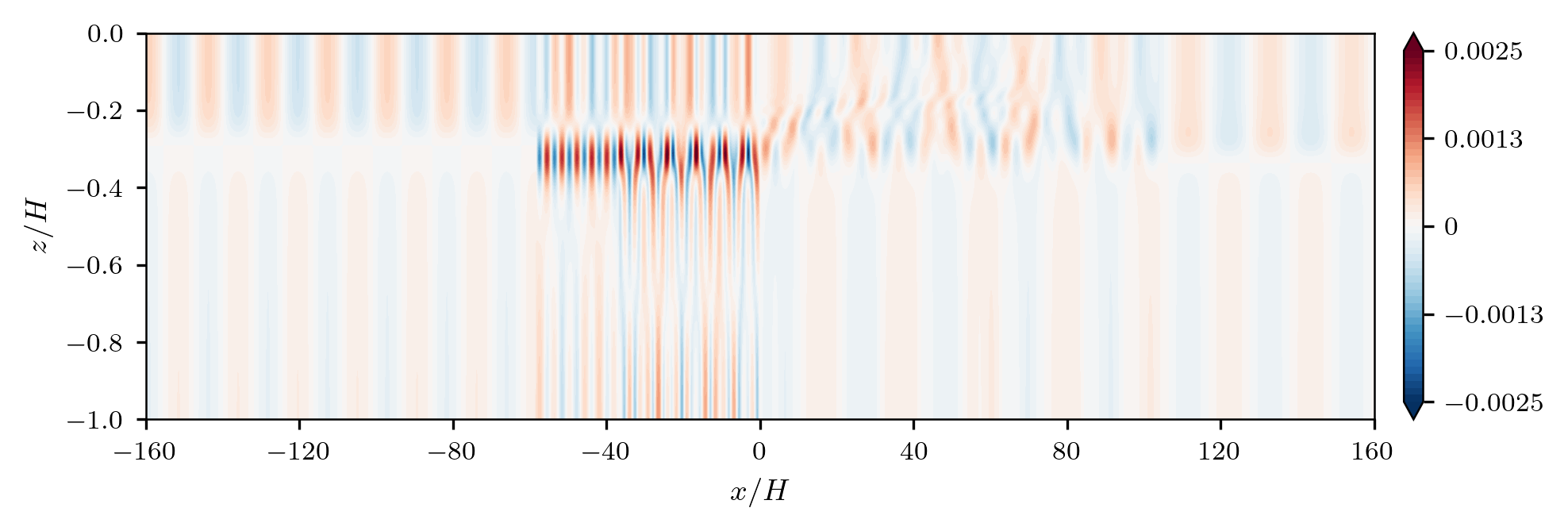}}\\
    \subfigure[ ]
              {\includegraphics[width=120mm]{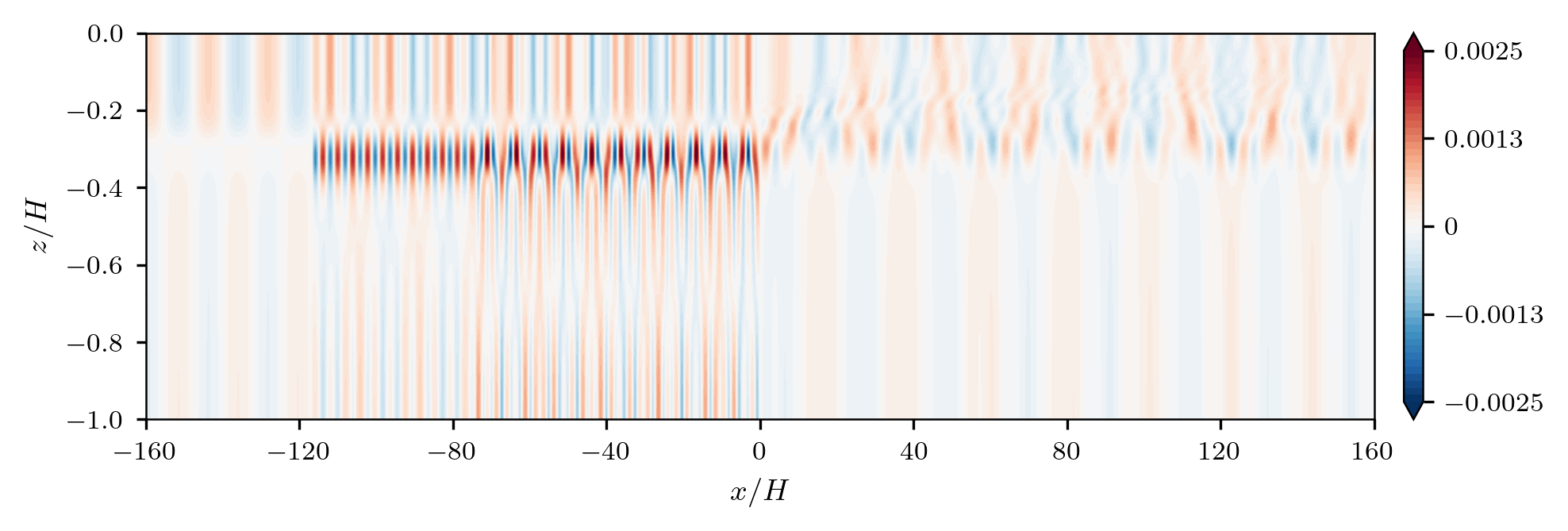}} 
  \caption{\label{f:fig7} Wave-induced dimensionless horizontal current fields $U/(NH)$  at (a) $t = 15\,T$ and (b) $t = 30\,T$ for the pycnocline case, for which $N(z)$ is shown in figure~\ref{fig:pyc_setup}. Other set-ups are the same as those in figure~\ref{f:fig1}. }
\end{center}
\end{minipage}
\end{center}
\end{figure}   
    Figure~\ref{f:fig7} shows the wave-induced dimensionless horizontal current fields for the pycnocline case.  
    Increasing the integration time from $t=15T$ (figure~\ref{f:fig7}a) to $t=30T$ (figure~\ref{f:fig7}b) leaves the near field essentially unchanged: the peak values                                                       
     $\max|u_{\mathrm{lin}}|=2.74\times10^{-3}$ and  $\max|u_{\mathrm{tot}}|=2.78\times10^{-3}$ (in units of $NH$), as well                                                      
     as the second-order bound contribution                                                                                      
     ($\max|\delta u|=1.13\times10^{-4}$, about $4.1\%$ of                                                                     
     $\max|u_{\mathrm{lin}}|$), are the same as at $t=15T$, indicating that                                                      
     the near field above the ridge has already saturated under the steady                                                       
     tidal forcing.  The wave                                                           
     trapped at the pycnocline depth ($z\simeq-0.3H$) stretches into a                                                           
     continuous band along $x$; and the rightward far field ($x>0$)                                                              
     develops a more complete beam structure.  The bounded contribution                                                          
     remains concentrated in the pycnocline layer and does not grow with                                                         
     the integration time.

     As shown in figure~\ref{f:fig7}, when the constant-$N$ background is replaced by the pycnocline profile (while with the depth‑averaged  $N^2$ kept fixed), the peak wave-induced velocity drops by about  $45\%$, from $4.97\times10^{-3}$ to $2.78\times10^{-3}$ (in units of $NH$) relative to figure~\ref{f:fig1}.    
     A mode-by-mode diagnosis  shows that this results  from several cooperating effects as follows.              
                                                                                                           
     \paragraph{(i) The primary cause is weaker projection of the tidal forcing onto the modes.}                                                                                                           
     The modal amplitude, $A_{mn}$ with $n=1$, is systematically smaller in the pycnocline case                                                                            
     (Table~\ref{tab:modal_2}).  Because $94\%$ of the water column now                                                     
     carries only $N_b^2 = 0.1\,N_{\mathrm{ref}}^2$, the effective                                                               
     stratification is weaker and the leftward phase speeds drop                                                                 
     (e.g.\ $c/(NH)=0.134\to0.087$ for $m=2$); the associated wavenumbers                                                        
     $|k|=\omega_0/|c|$ grow, while both the ridge transform $\hat{h}(k)$                                                        
     (for a Gaussian ridge, decaying exponentially with $|k|$) and the                                                           
     Bessel function $J_n(-kU_0/\omega_0)$ decrease rapidly with $|k|$.  The                                                       
     higher leftward modes are therefore almost entirely decoupled from the                                                      
     forcing ($|A|$ is $33\times$ smaller for $m=4$ and $\sim\!600\times$                                                        
     smaller for $m=5$).  Even for the gravest rightward mode, whose $k$                                                         
     becomes {smaller} (so that $\hat{h}$ increases), the denominator                                                       
     $k\,\partial\phi/\partial k$ in (\ref{eq:Anm_review}) grows by more, and $|A|$ is still reduced                                                      
     by a factor of $2.1$.                                                                                                       
                                                                                                                                 
     \begin{table}                                                                                                               
       \centering                                                                                                                                                                                                                                                                                                             
       \begin{tabular}{llccc}                                                                                                    
         direction & $m$ & $c/(NH)$ const-$N$ $\to$ pycn. &                                                                      
         $|A|$ (m\,s$^{-1}$) const-$N$ $\to$ pycn. & reduction \\                                                                
         \hline                                                                                                                  
         right & 1 & $0.361\to0.475$ & $1.79\times10^{-3}\to8.4\times10^{-4}$ & $2.1\times$ \\                                   
         right & 2 & $0.191\to0.155$ & $2.68\times10^{-3}\to3.8\times10^{-4}$ & $7\times$ \\                                     
         left  & 2 & $0.134\to0.087$ & $3.93\times10^{-3}\to1.1\times10^{-3}$ & $3.5\times$ \\                                   
         left  & 4 & $0.065\to0.040$ & $2.14\times10^{-3}\to6.4\times10^{-5}$ & $33\times$ \\                                    
         left  & 5 & $0.051\to0.029$ & $8.3\times10^{-4}\to1.4\times10^{-6}$ & ${\sim}600\times$ \\                              
       \end{tabular}        
       \caption{Phase speeds and modal forcing amplitudes of the first five                                                      
       $n=1$ modes: constant $N$ (figure~\ref{f:fig1}'s parameters, $d_s/H=0.08$) versus                                                      
       pycnocline $N(z)$ (with $d_s/H=0.06$).    \label{tab:modal_2}   }                                                                                                        
     \end{table}

     \paragraph{(ii) Partial compensation comes from trapping of the remaining energy in the pycnocline                                                          
     layer.}                                                                                              
     With the normalization $\phi'(-H)=1$, the pycnocline case's eigenfunctions                                                         
     reach $\max|\phi'| = 1.2$--$7.5$, compared with $\approx1.0$--$1.5$                                                         
     for constant $N$. As a consequence, the modal structure is strongly peaked at                                                                 
     $z \simeq z_s = -0.3H$.  Consistently, the maximum of the total field is                                                          
     found inside the pycnocline.                                                           
     This in-layer amplification of $\phi'$ (up to $7\times$) partially                                                          
     compensates the weaker wave amplitude, so the overall peak decreases by only                                                         
     $45\%$ rather than by an order of magnitude.

     In summary, confining the stratification to a layer of thickness  $0.06H$ weakens the projection of the tide--topography forcing onto the discrete modes---dramatically so for the higher leftward                                                                
     modes---and traps the residual wave energy inside the pycnocline.                                                           
     The peak of $|u|$ therefore drops to $2.78\times10^{-3}\,(NH)$ and                                                          
     moves from the surface down to $z\simeq-0.3H$.

\begin{figure}
\centering
\includegraphics[width=0.62\textwidth]{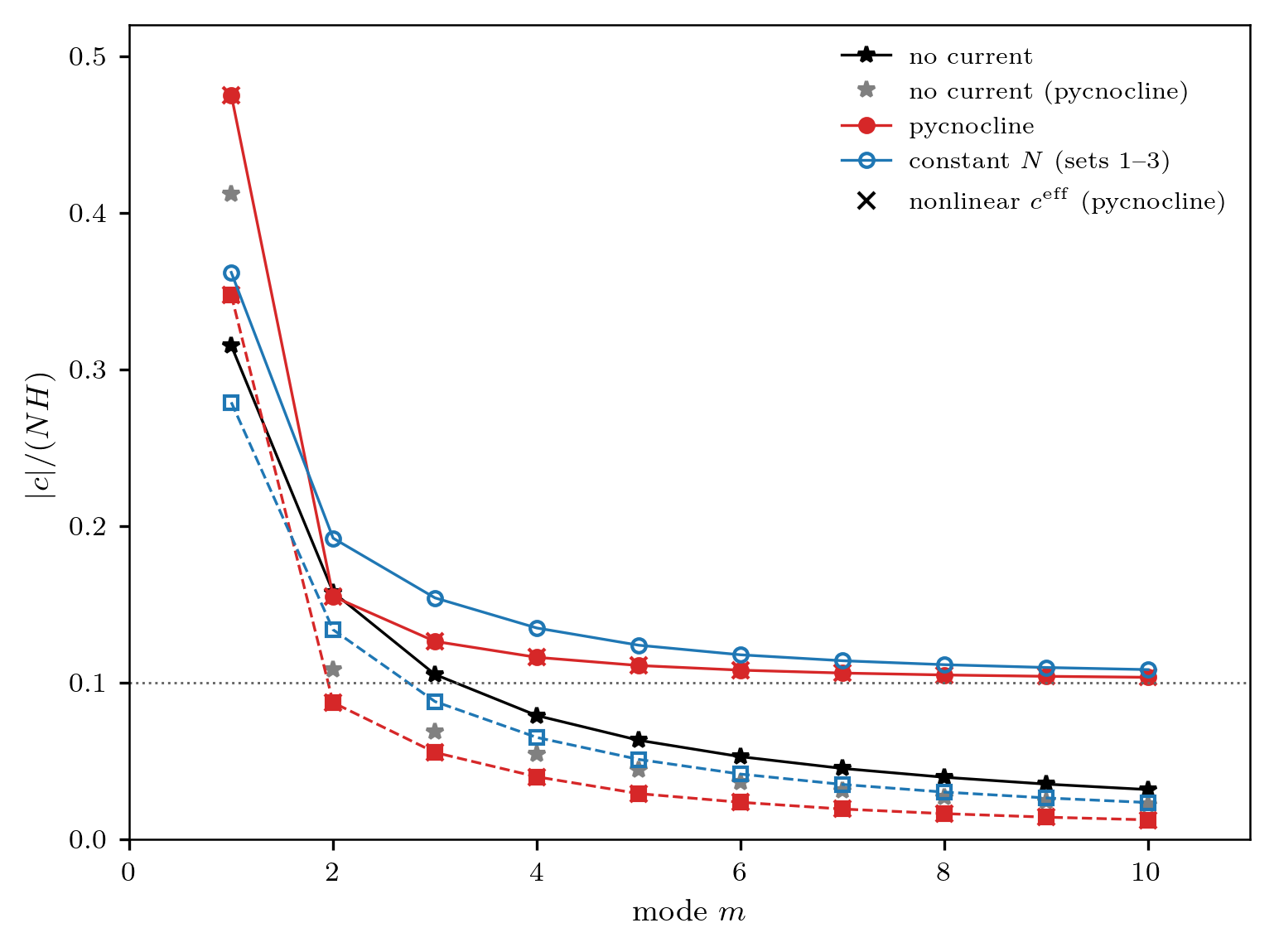}
\caption{Phase speeds $|c|/(NH)$ versus mode number for the
pycnocline case (\ref{eq:pycnocline}) (red, filled) versus the
constant-$N$ sets (blue, open): circles/solid rightward,
squares/dashed leftward.  Stars: no-current spectra for uniform $N$
(black) and for the pycnocline case (grey).  Crosses: third-order corrected
$c^{\rm eff}$ for the pycnocline case.  Dotted line:
$U_{s}/(NH)=0.1$.}
\label{fig:pyc_ps}
\end{figure}
Figure~\ref{fig:pyc_ps} shows the phase speeds.  The pycnocline
restructures the spectrum: the no-current gravest mode accelerates
from $0.315$ to $0.412\,NH$ (the stratification maximum near the lid
acts almost as an interface), and the rightward gravest mode reaches
$0.475\,NH$ versus $0.362\,NH$ for uniform $N$ ($N$ in the ordinate of figure~\ref{fig:pyc_ps} 
and in the above normalisation is also constant).

Compared to the uniform-$N$ case, the leftward
wavenumbers increase by a factor $\simeq1.5$ from mode~2 up, with two
consequences: the ridge spectrum $\hat h\propto\ee^{-(ak)^2/4}$ now
cuts the leftward amplitudes off beyond mode~4 ($|A|$ drops by six
orders of magnitude from mode~4 to mode~10), so the upstream tide
collapses onto modes 1--2; and these surviving modes carry their
displacement maxima inside the pycnocline.  On the other hand, the third-order corrected
$c^{\rm eff}$ remains invisible at the scale of figure~\ref{fig:pyc_ps}.

Figure~\ref{fig:pyc_dk} shows that the nonlinear corrections, however, are strongly amplified. The wavenumber correction peaks at $|\delta k|\,H=2.98\times10^{-3}$ (leftward mode~2) --- a factor
$\sim67$ larger than the constant-$N$ maximum $4.43\times10^{-5}$. Two structural features accompany the growth.
First, the sign of $\delta k/k$ remains uniformly positive for all ten modes of both directions: the pycnocline amplifies the magnitude of the nonlinear slow-down without reversing its sign.  Second, the modal decay of the
correction now tracks the pycnocline-restructured amplitudes: the leftward correction collapses beyond mode~4, while the rightward correction persists across all ten modes that have been calculated.

\begin{figure}
\centering
\includegraphics[width=0.72\textwidth]{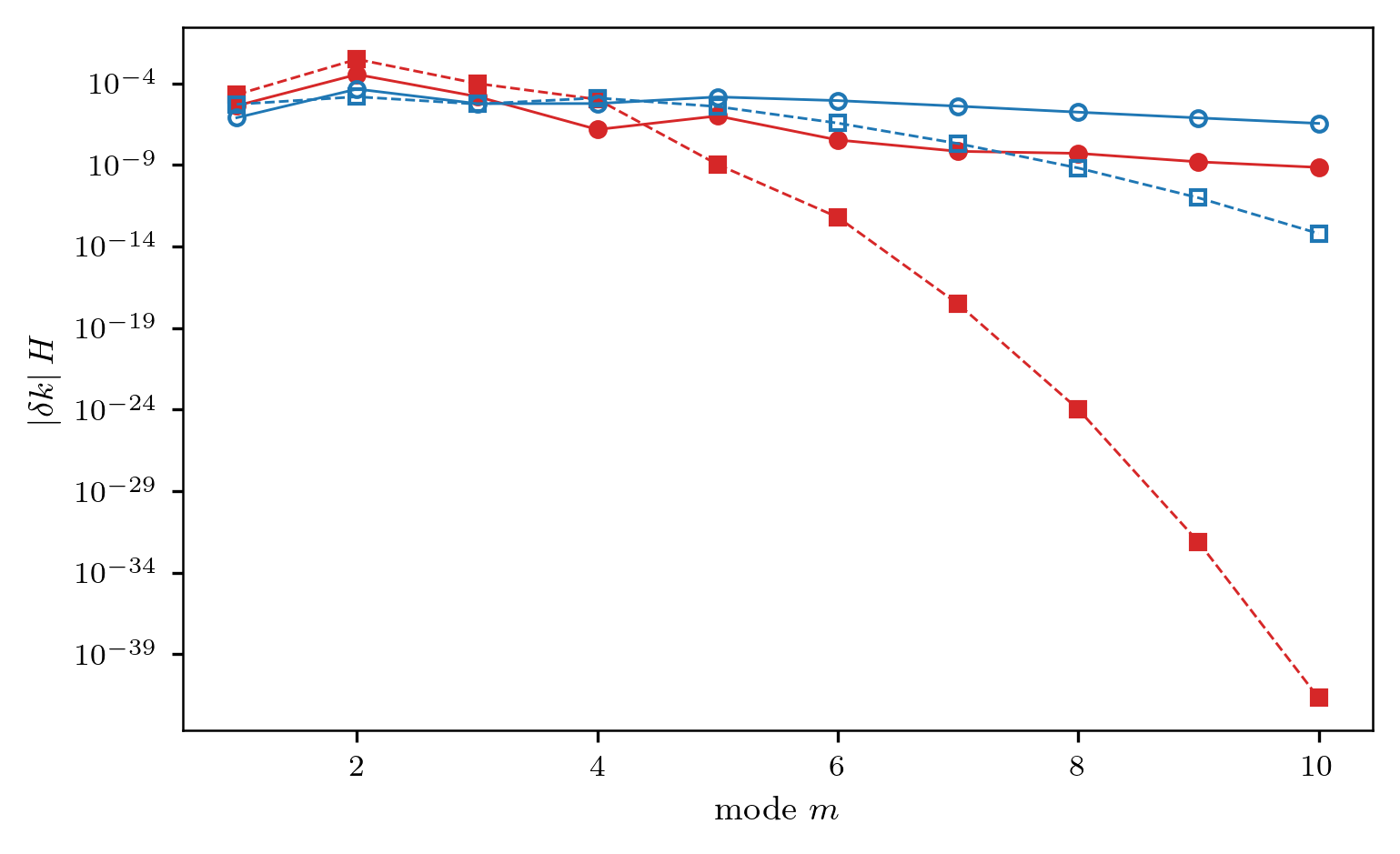}
\caption{Wavenumber correction $|\delta k|\,H$ versus mode number for
the pycnocline case (red, filled) versus constant-$N$ sets~1--3 (blue,
open): circles/solid rightward, squares/dashed leftward.  The
pycnocline amplifies the peak correction by a factor $\sim67$
(leftward mode~2), without reversing its sign (see text).}
\label{fig:pyc_dk}
\end{figure}
{The waveform distortion also grows correspondingly}
(figure~\ref{fig:pyc_wave}, to be compared panel by panel with the
constant-$N$ figure~\ref{fig:wave}): the bound second
harmonic reaches $8.1\%$ of the fundamental for rightward mode~2 and
$5.9\%$ for leftward mode~2 (uniform $N$: at most $2.3\%$).  The
structure of the distortion changes as much as its magnitude.  In the
constant-$N$ case the bound harmonic of the rightward modes peaks just
inside the surface shear layer ($z/H\simeq-0.15$) while the leftward
distortion follows the modal displacement at depth; here the
eigenfunctions are compressed into the pycnocline, so $\eta_1$ and
$\eta_2$ alike are trapped in the layer $z/H\simeq-0.2$ to $-0.4$, and
the local wave slope $k\max|\eta|$ is correspondingly larger.  The
directional contrasts of figure~\ref{fig:wave} persist: the higher
leftward modes are essentially unforced beyond mode~4, whereas the
rightward modes, approaching the spectral accumulation point, retain
finite amplitudes out to mode~8.  The net picture is that
concentrating the stratification where the shear lives makes the wave
field \emph{markedly more nonlinear} --- larger $\delta k$, larger distortion
--- whereas the mean state becomes \emph{less} shear-unstable.  At the
parameters of \citet{LambDunphy2018} the field remains comfortably
inside the weakly nonlinear regime, but the pycnocline case shows that the nonlinear corrections are
sensitive to where $N^2$ sits relative to $\Ue'$.
\begin{figure}
\centering
\includegraphics[width=0.98\textwidth]{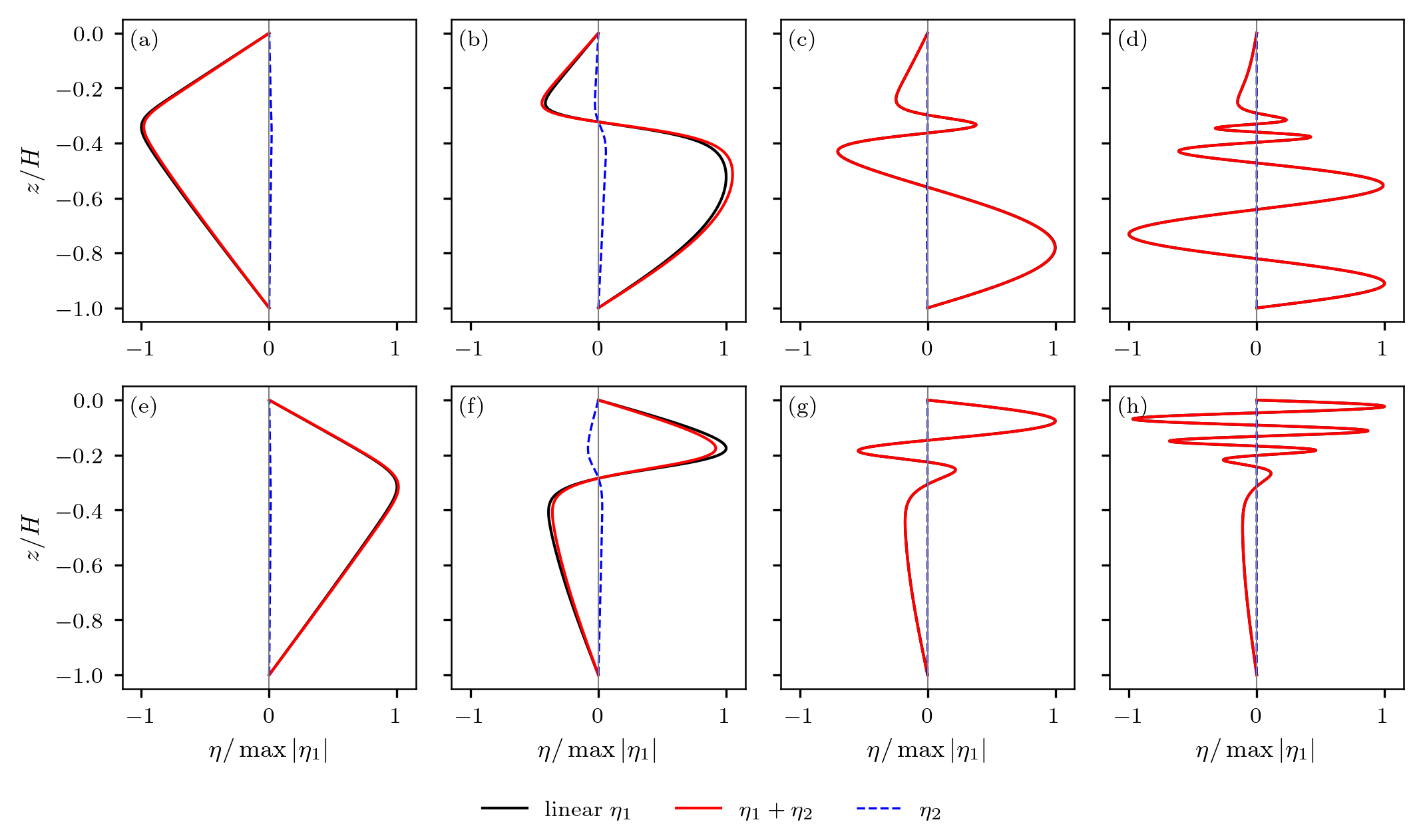}
\caption{Displacement profiles $\eta(z)=b/N^2$ at the crest phase for
the pycnocline case, in the same layout and normalisation as
figure~\ref{fig:wave}: (a)--(d) leftward modes $m=1,2,4,8$; (e)--(h)
rightward modes $m=1,2,4,8$.  Black: linear first-order profile
$\eta_1$; blue dashed: bound second harmonic $\eta_2$; red: total.
Per-mode data ($c$ in m\,s$^{-1}$; $\delta k/k$; distortion ratio
$\max|\eta_2|/\max|\eta_1|$; $\max|\eta_1|$ in m): leftward $m=1$:
$-1.74$, $+5.0\times10^{-5}$, $1.9\times10^{-2}$, $1.76$; $m=2$:
$-0.44$, $+1.8\times10^{-3}$, $5.9\times10^{-2}$, $6.62$; $m=4$:
$-0.20$, $+2.9\times10^{-6}$, $4.0\times10^{-3}$, $0.86$; $m=8$:
$-0.08$, $\simeq0$, $1.9\times10^{-9}$, $4.2\times10^{-10}$
(essentially unforced); rightward $m=1$:
$+2.38$, $+1.4\times10^{-5}$, $1.2\times10^{-2}$, $1.17$; $m=2$:
$+0.77$, $+3.8\times10^{-4}$, $8.1\times10^{-2}$, $1.82$; $m=4$:
$+0.58$, $+1.2\times10^{-7}$, $1.0\times10^{-3}$, $1.46$; $m=8$:
$+0.52$, $+3.8\times10^{-9}$, $8.8\times10^{-4}$, $0.48$.}
\label{fig:pyc_wave}
\end{figure}

\subsection{The increased nonlinearity at the increased Richardson number}
We have shown that the nonlinear corrections grow by a factor of $\sim 67$, while the minimum Richardson number rises from $1.03$ to $6.5$. One may wonder why does nonlinearity strengthen for the pycnocline case when the Richardson number increases?
This is a subtle but important point. The intuition that ``larger Richardson number $\Rightarrow$ more stable background $\Rightarrow$ weaker nonlinearity'' is correct for the {mean shear flow itself}, but it does {not} apply to the {wave-induced nonlinearity}. The strength of the nonlinear corrections is not governed by the background Richardson number; however, it is governed by {how strongly the wave field overlaps with the gradients of the background state}, and by {how large those local gradients are}.

More specifically,  Richardson number $Ri$ measures background shear stability, not wave--shear coupling strength.
For the pycnocline case, when $N^2(z)$ is co-located with the shear layer, the density gradient becomes very strong. Although the velocity gradient $\bar U'$ is also large, the increase in $N(z)^2$ at certain depth is much larger than the increase in $\bar U'^2$. Since $Ri(z) = N^2(z)/(\bar U')^2$, the background flow itself becomes more stable against Kelvin--Helmholtz instability. 
On the other hand, the Thorpe-type nonlinearity is driven by products of the wave-induced velocity shear and the background shear, i.e. $\bar U' \cdot \phi'$, and by the vertical gradients of the wave field. When $N^2(z)$ and $\bar U'$ are co-located, the eigenfunction $\phi(z)$ is strongly trapped and compressed into this narrow region (the pycnocline/shear layer). This localisation dramatically amplifies the wave strains and the nonlinear forcing.

Second, the nonlinear kernels depend directly on the co-located gradients. For the convenience of readers, 
the diagonal kernel (3.4), which determines the second-order forcing $S_2$, is repeated below: 
\begin{equation}
\nonumber \mathcal{R}_{pp} = kM'(z)\phi^2,\qquad 
\mathcal{S}_{pp} = \frac{k\left[N^2(z)\bar U'(z) - (N^2)'(z)(\bar U - c)\right]}{(\bar U - c)^2}\phi^2.
\end{equation}
\noindent When $N^2$ and $\bar U'$ are co-located, both $M'(z)$ (the vertical derivative of the Taylor--Goldstein coefficient) and the term $N^2\bar U'$ become very large at those depths. This produces an exceptionally strong nonlinear forcing source inside the pycnocline. Even though $Ri$ is high (strong stratification), the wave eigenfunction $\phi$ passes through this region and experiences an intense vertical strain ($\partial \eta/\partial z$), generating second harmonics and wavenumber corrections far larger than in the uniform-$N$ case.

Third, modal structure is much focused (eigenfunction compression).
In a uniform stratification (constant $N$), eigenfunctions are smooth sinusoidal profiles across the whole depth. In a strong pycnocline, wave energy is localised near the maximum density gradient (i.e. the shear-layer centre). This leads to much larger local displacements $\eta$ and velocities $u$ at that depth than in the uniform-$N$ case, even if the global modal amplitude $A_{nm}$ is the same. Because the wave is compressed into a thinner layer, the local wave slope $k\max|\eta|$ increases significantly. This is why the distortion ratio $\eta_2/\eta_1$ jumps from $2.3\%$ to $8.1\%$.

Finally, the phase-speed change modifies the wavenumber matching. Figure~\ref{fig:pyc_ps} shows that the pycnocline raises the no-current phase speed from $0.315$ to $0.412\,NH$. This structural change alters the denominator $(\nu - \kappa\bar U)$ in the second-order forcing (\ref{eq:S2diag}), shifting the near-resonance conditions. Such a shift directly amplifies the magnitude of $\delta k$.

In summary, a more stable background with larger $Ri$ does not necessarily suppress wave nonlinearity. On the contrary, when stratification and shear are co-located, the wave energy and strain are focused into a narrow region, greatly enhancing the nonlinear kernels ($N^2\bar U'$ and $M'$). As a result, the nonlinear wavenumber correction and waveform distortion become orders of magnitude larger than in the uniform-$N$ case -- exactly as observed in \S \ref{sec:pycnocline}.

\section{Open problems}
\label{sec:outlook}

While the theoretical framework developed in \S\ref{sec:higher-order} is general, the numerical results of \S\ref{sec:results} are restricted to the diagonal self-interaction of individual modes. Several important extensions remain, which we outline below in roughly decreasing order of anticipated physical significance.

\begin{enumerate}
\item \textit{Off-diagonal triads and the $n=0$ mean field.}  The
general mode-pair kernels (\ref{eq:knabfix})--(\ref{eq:kbfix}) and the
residue formula (\ref{eq:A2}) cover all pairs $(p,q)$, but only $p=q$
has been evaluated.  Near-resonant pairs (detuning length
$\ell_{\rm det}$, \S\,\ref{sec:triad}) could focus energy into selected
modes and amplify their nonlinear corrections well beyond the
single-mode diagonal estimates of \S\,\ref{sec:results}.  On the other hand, the $n=0$ mean field
(\S\,\ref{sec:meanfield}) is singular where $\Ue\to0$ and requires a
critical-layer regularisation. 

\item \textit{Boundary data $g_2$ and $\hat g_3$.}  The
$O(\varepsilon^2)$ bottom-boundary forcing (\ref{eq:g2})--(\ref{eq:B4})
and its third-order counterpart were derived but not computed.  The
$g_2$ term drives a free, Lamb-type response at $2n\omega_0$ that
radiates independently of the bound harmonic; whether it contributes
materially to the superharmonic tide observed in numerical simulations
\citep{GayenSarkar2011} is an open question.

\item \textit{Continuous spectrum.}  The discrete-spectrum truncation
omits the continuous-spectrum (critical-layer) contribution that
accumulates near $c\to\Ue^{\max}$.  Recent work on the continuous
spectrum of sheared internal tides \citep{OnukiVenaille2025} shows
wave-packet shear growing secularly along characteristics; coupling
this contribution into the discrete-spectrum expansion would complete
the picture near the accumulation
point, where our discrete modes already show the largest sensitivity.

\item \textit{Stronger currents.}  All diagnostics degrade with
$U_s/(NH)$; the set-8 case's currents of \citet{LambDunphy2018}
($U_s/(NH)\gtrsim0.2$) should push the Richardson-number diagnostic
toward $1/4$ first and bring the weakly nonlinear corrections into a
regime where they are directly visible in simulations.  

\item \textit{Three-dimensional effects.}  The expansion conducted in this study is strictly
two-dimensional.  Shear-induced breaking of internal waves is known to
develop spanwise (convective-roll) instabilities well before the
two-dimensional overturning threshold \citep{Howland2021}; a
three-dimensional stability analysis of the finite-amplitude modes
computed here would test whether the same holds for the forced tide.

\item \textit{Non-uniform stratification.}  The pycnocline case of
\S\ref{sec:pycnocline} is a first step: it shows that co-locating
$N^2$ with $\Ue'$ amplifies the nonlinear corrections by up to two
orders of magnitude.  What remains is to
map the parameter space systematically --- pycnocline offset from the
shear layer, and realistic oceanic $N(z)$ --- and
to revisit the eigenvalue-scan ranges, which the restructured spectrum
can silently outgrow.

\item \textit{Parametric subharmonic instability.}  PSI does not lie on
the integer harmonic grid of the forced expansion
(\S\,\ref{sec:triad}); it requires a multiple-scale,
slowly-varying-amplitude description of the radiated beams and is
outside the successive-approximation framework adopted in this study.

\item \textit{Validation against simulations.}  The predicted bound
second harmonic (0.3--3\% of the fundamental at the parameters of
\citet{LambDunphy2018}) and its shear-layer localisation are directly
testable against nonlinear numerical simulations of tidally forced
flow over ridges in the spirit of \citet{GayenSarkar2011} and
\citet{KlemaVenayagamoorthy2024}, whose superharmonic fields reach the
10--15\% level at stronger forcing.  Such a comparison would also
calibrate where the weakly nonlinear ordering ceases to be
quantitative.
\end{enumerate}

\section{Conclusions}
\label{sec:conclusions}

We have embedded \citet{Thorpe1978}'s successive-approximation expansion into the
discrete-spectrum generation theory of \citet{LambDunphy2018} and
quantified the leading finite-amplitude corrections.  The linear
discrete-spectrum model is excellent at these parameters: nonlinear
dispersion shifts the wavenumbers by at most
$|\delta k|\,H\sim1.2\times10^{-4}$ and waveform distortion reaches at
most a few per cent.  The corrections are
structured: they peak at modes 2--3, grow quadratically with ridge
height and with the strength of the surface current, decrease the phase
speed at fixed frequency (a nonlinear slow-down), and are concentrated in the surface
shear layer for downstream-propagating modes.  This explains, from within the
theory itself, why the linear discrete-spectrum model compared so well with fully nonlinear
simulations for the present configurations.

The new pycnocline case (\S\ref{sec:pycnocline}) qualifies
this verdict in an important way: with the stratification maximum
co-located with the shear layer, the nonlinear wavenumber correction
grows by a factor $\sim67$ and the waveform distortion reaches
$8\%$ --- all while
the background Richardson number rises.  The uniform-stratification
conclusion of robust linear theory is therefore not generic: it
depends on where $N^2$ sits relative to $\Ue'$, and oceanic
pycnocline stratification moves the forced tide measurably closer to
the finite-amplitude regime.

Beyond the quantitative verdict, the calculation delivers the machinery
for genuinely finite-amplitude studies of tidally forced flow over
topography: the mode-pair kernels
(\ref{eq:knabfix})--(\ref{eq:kbfix}) with their diagonal reduction
(\ref{eq:diag}), the forced Taylor--Goldstein problem (\ref{eq:S2}) for
bound harmonics, and the solvability-based wavenumber correction
(\ref{eq:delta_k_final}) with the explicit Long-equation resonant
projection (\ref{eq:long})--(\ref{eq:dklong}), validated in the
free-wavetrain limit against \citet{Thorpe1978} (Appendix~\ref{app:thorpe}).  The extensions listed in \S\,\ref{sec:outlook} ---
off-diagonal triads, the mean field, boundary data, continuous
spectrum, stronger currents, and three-dimensional effects --- define
the path from the present weakly nonlinear corrections toward a
quantitative finite-amplitude theory of topographic internal tides.

\section*{Acknowledgements}
This work is partly supported by National Science Foundation of China (Grant No. 12432016 \& No. 12272007).

\appendix
\section{Validation against \citet{LambDunphy2018}}
\label{app:lamb}

The first-order solution of \S\,\ref{sec:first-order-review} is the
linear discrete-spectrum model of \citet{LambDunphy2018}; its numerical
implementation (shooting solution of the Taylor--Goldstein equation
(\ref{eq:Ln}), eigenvalue scan, and residue amplitudes
(\ref{eq:Anm_review})) was validated by reproducing their published
results at their own parameters.  Figure~\ref{fig:ps} of the main text
reproduces their figure~5 (phase speeds of the first ten vertical modes), and figure~\ref{fig:lamb6}
reproduces their figure~6 (the eigenfunctions of modes 1, 2, 4 and 8); the $t=15\,T$ field of
figure~\ref{f:fig1} follows the format of their figure~1($b$).  In all cases, the results are visually indistinguishable from the published data.
\begin{figure}
\centering
\includegraphics[width=\textwidth]{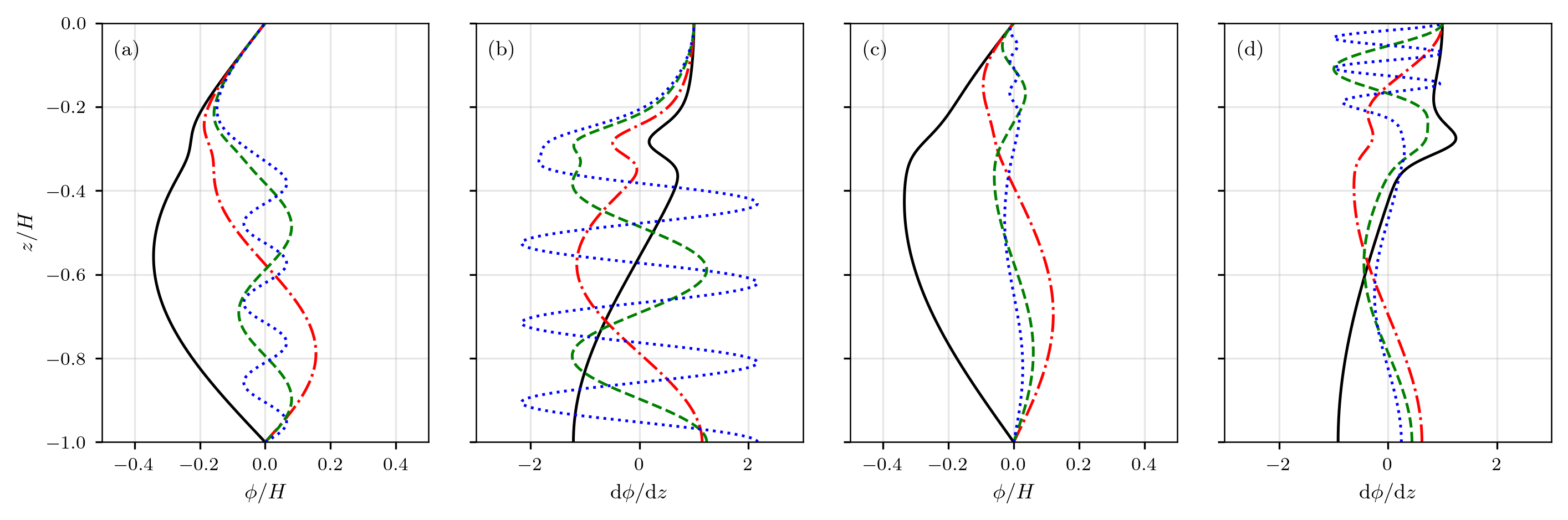}
\caption{Reproduction of figure~6 of \citet{LambDunphy2018}: vertical
structure of the first-order eigenmodes $m=1,2,4,8$ for the sets~1--3
current, $(U_s/(NH),z_s/H,d_s/H)=(0.1,-0.3,0.06)$, normalised to
$\phi(0)=0$, $\phi'(0)=1$ as in the original.  (a,b) leftward modes:
(a) $\phi/H$; (b) $\dd\phi/\dd z$.  (c,d) rightward modes: (c)
$\phi/H$; (d) $\dd\phi/\dd z$.  Black solid: mode~1; red dash-dotted:
mode~2; green dashed: mode~4; blue dotted: mode~8.}
\label{fig:lamb6}
\end{figure}

\section{Validation against \citet{Thorpe1978}}
\label{app:thorpe}

Reducing the forced formulation of \S\,\ref{sec:higher-order} to a
single mode without topography, whose frequency is free to adjust,
recovers the free-wavetrain problem of \citet{Thorpe1978}; this limit
provides an independent validation of both the derivation and the
numerical implementation.  On \citet{Thorpe1978}'s tanh profiles ($\beta h=20$ and
$30$) we have reproduced his figures~1--3, including its $U=0$
 limit.  The comparison (figures~\ref{fig:thorpe1}--\ref{fig:thorpe3})
confirms his linear and second-order results: the phase and group
speeds, the profiles of $\psi_1,\psi_2,\psi_3,\eta_1,\eta_2$ at all
seven shears, and the extremum depths of $\eta_1$ and $\eta_2$ are all
recovered.  
\begin{figure}
\centering
\includegraphics[width=0.7\textwidth]{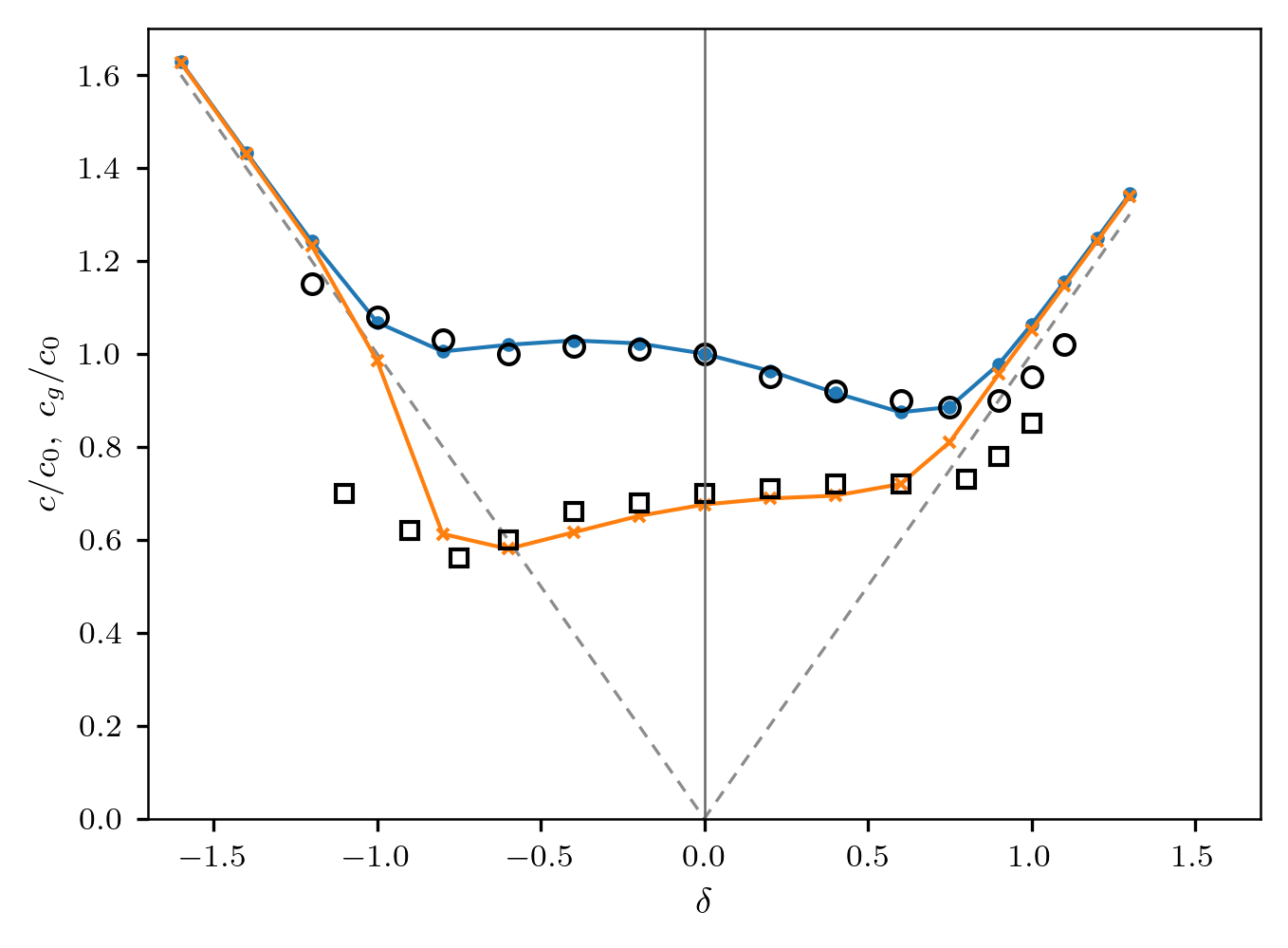}
\caption{Phase speed $c/c_0$ and group speed $c_g/c_0$ of the first
mode versus shear parameter $\delta$, $\beta h=20$.  Blue solid curve
with small dots: our $c/c_0$; orange solid curve with small crosses:
our $c_g/c_0$; black open circles/squares: digitised readings of
figure~1 of \citet{Thorpe1978} ($c$, $c_g$); grey dashed lines: the
range of the background velocity $U$.  Our linear Taylor--Goldstein
solution reproduces his figure~1 point by point ($c_0=0.457$ at
$\delta=0$).}
\label{fig:thorpe1}
\end{figure}
\begin{figure}
\centering
\includegraphics[width=\textwidth]{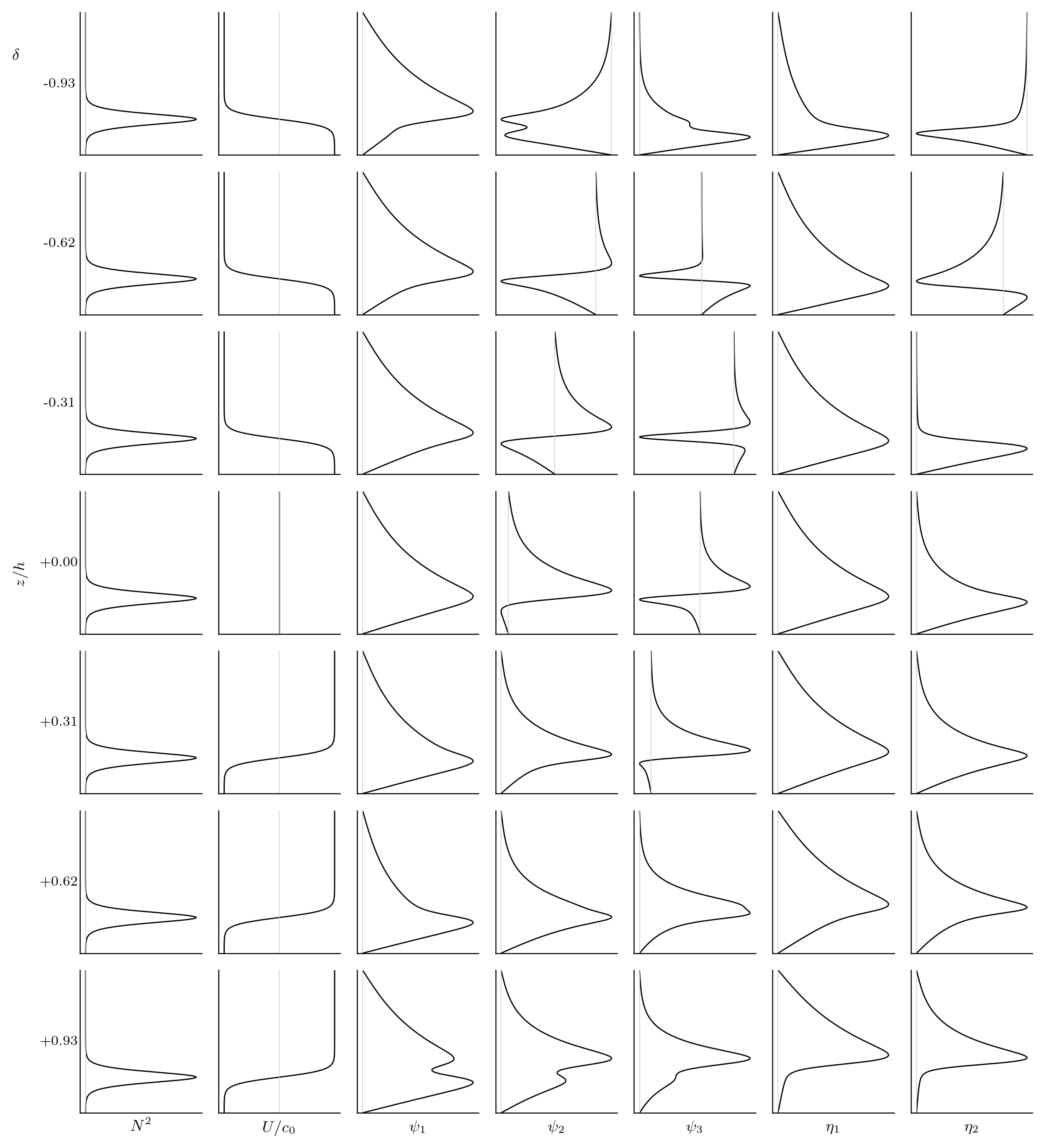}
\caption{Reproduction of figure~2 of \citet{Thorpe1978} ($\beta h=20$)
in his original layout: columns $N^2$, $U/c_0$, $\psi_1$, $\psi_2$,
$\psi_3$, $\eta_1$, $\eta_2$; rows
$\delta=-0.93,\ldots,+0.93$.  All profiles are black solid curves,
panel by panel corresponding to his figure~2.  Profile shapes,
extremum levels and signs agree at all seven shears.}
\label{fig:thorpe2}
\end{figure}
\begin{figure}
\centering
\includegraphics[width=0.95\textwidth]{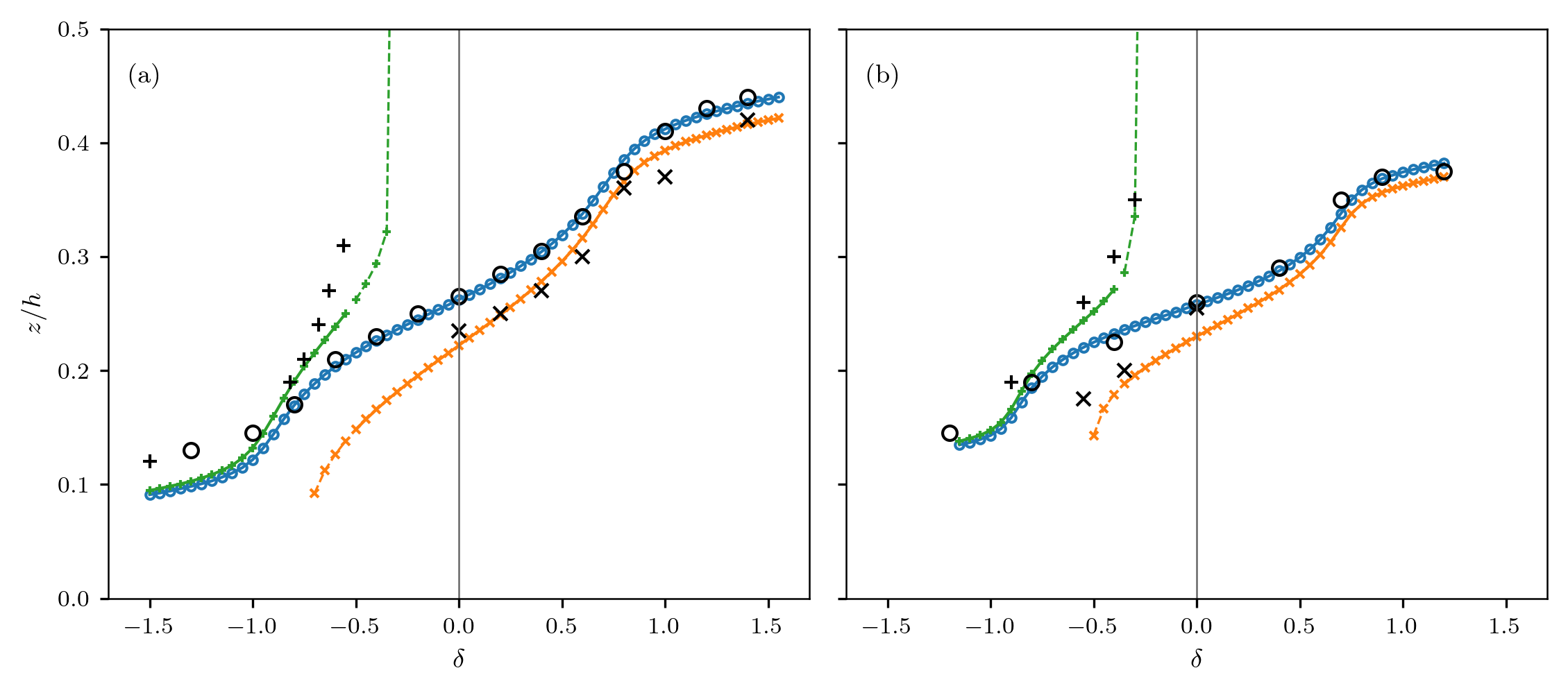}
\caption{Depths of the interior extrema of $\eta_1$ and $\eta_2$
versus $\delta$: (a) $\beta h=20$; (b) $\beta h=30$.  Blue solid with
open circles: $\max\eta_1$; orange solid with crosses: $\max\eta_2$;
green solid with plus signs: $\min\eta_2$ (segments where $|\eta_2|$
is not the global maximum shown dashed in the same colour); black open
circles/crosses/plus signs: digitised readings of figure~3 of
\citet{Thorpe1978}.  The extremum depths agree within read-off
accuracy (rms $\lesssim0.03h$), including the upturn of the
$\min\eta_2$ branch on the negative-$\delta$ side.}
\label{fig:thorpe3}
\end{figure}


\end{document}